\documentclass[twocolumn,tighten]{aastex63}

\usepackage{color}
\usepackage{comment}
\usepackage{amsmath}
\usepackage{amssymb}
\usepackage{tabularx}
\usepackage{nameref}
\usepackage{varioref}
\usepackage{hyperref}
\usepackage{cleveref}
\usepackage{xcolor}
\usepackage{soul}
\usepackage{rotating}
\usepackage{dcolumn}   
\usepackage{longtable}
\usepackage{enumitem}
\usepackage{scalerel}

\usepackage{soul}

\newcommand{\Dennis}[1]{{\bf\color{red} DZ: #1}}

\newcommand{\ra}[3]{$#1^\mathrm{h} #2^\mathrm{m} #3^\mathrm{s}$}
\newcommand{\dec}[3]{$#1^\circ #2' #3''$}

\newcommand{\note}[1]{\tablenotemark{\scaleto{\text{#1}}{4pt}}}
  
\shorttitle{UDG Properties Across Environment}
\shortauthors{Kadowaki et al.}

\begin{document}

\title{On the Properties of Spectroscopically-Confirmed Ultra-Diffuse Galaxies Across Environment}


\author[0000-0002-3767-9681]{Jennifer Kadowaki}
\affil{Department of Astronomy/Steward Observatory, 933 North Cherry Avenue, Rm. N204, Tucson, AZ 85721-0065, USA}
\email{jkadowaki@email.arizona.edu}

\author[0000-0002-5177-727X]{Dennis Zaritsky}
\affil{Department of Astronomy/Steward Observatory, 933 North Cherry Avenue, Rm. N204, Tucson, AZ 85721-0065, USA}

\author[0000-0001-7618-8212]{R. L. Donnerstein}
\affil{Department of Astronomy/Steward Observatory, 933 North Cherry Avenue, Rm. N204, Tucson, AZ 85721-0065, USA}

\author[0000-0003-3714-2574]{Pranjal RS}
\affil{Department of Astronomy/Steward Observatory, 933 North Cherry Avenue, Rm. N204, Tucson, AZ 85721-0065, USA}

\author[0000-0001-8855-3635]{Ananthan Karunakaran}
\affiliation{Department of Physics, Engineering Physics and Astronomy Queen's University Kingston, ON K7L 3N6, Canada}

\author[0000-0002-0956-7949]{Kristine Spekkens}
\affiliation{Department of Physics, Engineering Physics and Astronomy Queen's University Kingston, ON K7L 3N6, Canada}
\affiliation{Department of Physics and Space Science Royal Military College of Canada P.O. Box 17000, Station Forces Kingston, ON K7K 7B4, Canada}


\begin{abstract}
We present new redshift measurements for 19 candidate, ultra-diffuse galaxies (UDGs) from the Systematically Measuring Ultra-Diffuse Galaxies (SMUDGes) survey after conducting a long-slit, spectroscopic follow-up campaign on 23 candidates at the Large Binocular Telescope. We combine these results with redshift measurements from other sources for 29 SMUDGes and 20 non-SMUDGes candidate UDGs. Together, this sample yields 44 spectroscopically-confirmed UDGs ($r_e\geq1.5$ kpc and $\mu_g(0)\geq24$ mag arcsec$^{-2}$ within uncertainties) and spans cluster and field environments, with all but one  projected on the Coma cluster and environs. 
We find no statistically significant differences in the structural parameters of cluster and non-cluster confirmed UDGs, although there are hints of differences among the axis ratio distributions. Similarly, we find no significant structural differences among those in locally dense or sparse environments. However, we observe a significant difference in color with respect to projected cluster-centric radius, confirming trends  observed previously in statistical UDG samples. This trend strengthens further when considering whether UDGs reside in either cluster or locally dense environments, suggesting starkly different star formation histories for UDGs residing in high and low-density environments.
Of the 16 large ($r_e \geq 3.5$ kpc) UDGs in our sample, only one is a field galaxy that falls near the early-type galaxy red sequence. No other field UDGs found in low density environments fall near the red sequence. This finding, in combination with our detection of {\sl GALEX} NUV flux in nearly half of the UDGs in sparse environments, suggest that field UDGs are a population of slowly evolving galaxies.
\end{abstract}


\section{Introduction}

Ultra-diffuse galaxies (UDGs) are simply defined as spatially-extended (typically requiring that the effective radius, $r_e$, be $\geq$1.5 kpc), low surface brightness galaxies (typically requiring that the optical central surface brightness, $\mu_g(0)$, be $\geq$24 mag arcsec$^{-2}$).
Such galaxies have been recognized for over 40 years \citep{Disney, Sandage1984, Vigroux1986, Impey1988, Bothun1991, McGaugh, Schwartzenberg1995, Dalcanton1997, Sprayberry1997, Conselice2003, Penny2008}. Work continued in the interim in trying to understand how these galaxies fit into an evolutionary picture \citep[e.g.,][]{Conselice2003, Sabatini2005, Penny2009, Penny2011, Penny2014}. Improvements in imaging technology and data reduction techniques  \citep[e.g.,][]{Mihos2005, Mihos2013, Slater2009, Abraham2014, vanDokkum2014, Zaritsky19, Infante} 
have recently reinvigorated the field with the discovery of large numbers of such galaxies in galaxy clusters and groups \citep{vanDokkum2015a, Koda2015, Mihos2015, Yagi2016, vanderburg2016, Mihos2017, Roman2017a, greco, Danieli2019, tanog}.

Our understanding of this class of galaxy is hampered by the lack of distance measurements and the broad selection criteria for UDGs, which lead to heterogeneous samples \citep{Martin2016, Merritt2016, Martinez-Delgado2016, Roman2017a, vanderburg2017, Leisman2017, Prole2019, Torrealba2019}. Distances are required to constrain the local environment and determine physical parameters, including the size of UDGs, which correlates with the enclosed mass (see Appendix \ref{appendix}). UDG mass measurements are necessary to disentangle the myriad of proposed UDG formation and evolutionary models. While previous studies have found evidence for the majority of UDGs to be low-mass galaxies \citep{Amorisco2016, Sifon2018} which suggests a substantial overlap between UDGs and low mass cluster galaxies identified in \cite{Conselice2003}, UDGs exhibit a wide range of properties. For example, internal dynamics measurements for DF44 show a halo mass of $1.6 \times 10^{11} \, M_\odot$ \citep{vanDokkum2019b}, which is comparable to the halo mass of the Large Magellanic Cloud \citep[$1.4 \times 10^{11} \, M_\odot$;][]{erkal}.
Although internal dynamical measurements for UDGs remain rare, the inferred halo masses for  UDGs with $r_e > 3$ kpc are typically $\gtrsim 10^{11}$ M$_\odot$ (Appendix A). Such findings suggest that a fraction of UDGs live in relatively massive dark matter halos ($M>10^{11} \, M_\odot$). The origins of this subset of UDGs may well be very different from their less massive counterparts, which include cluster dwarf ellipticals (dEs) and dwarf spheroidals (dSphs) studied extensively by \citep{Sandage1984, Conselice2003, Penny2009}. The SMUDGes Survey catalogs this subset of potentially large (and therefore massive) UDGs \citep{Zaritsky19}.

Two inferences drawn from recent results where investigators were able to estimate the total mass for small, select samples of UDGs
drive the current flurry of interest in UDGs. First, at least some UDGs appear to be examples of galaxies with highly inefficient star formation or ``failed galaxies" \citep{vanDokkum2015a}. The extremely large effective radii ($> 4$ kpc) of some UDGs and their survival against the tidal forces present in the dense cluster environment together suggest that these may reside within large dark matter halos. Given their
low stellar luminosities,  these galaxies may be up to a hundred times less efficient at forming stars than an L$^*$ galaxy.  Because this result was first obtained for galaxies in the Coma galaxy cluster \citep{vanDokkum2015a,Koda2015} and there is evidence for bluer galaxies outside of clusters \citep{Roman2017a,Roman2017b}, it was broadly speculated that UDGs were ``failed'' galaxies due to  environmental processes that removed the gas early and effectively \citep[e.g.,][]{vanDokkum2015a,Roman2017b}. Given the extreme properties of UDGs, the environmental differences among UDGs might be even greater than those among high surface brightness galaxies. However, the situation may be more complicated than this simple scenario suggests because although environmental color differences have been found \citep{Roman2017a, greco, tanog},
low star formation efficiencies are observed even for HI-bearing field UDGs \citep{Leisman2017}.

Second, at least some UDGs appear to be the most massive examples of galaxies that are dark matter dominated even within their optical radii. As such, UDGs provide a probe of dark matter halo profiles that are minimally disturbed by baryonic processes. 
Dynamical measurements of the mass of UDGs can be determined using either the unresolved stellar light \citep{vanDokkum2016} or globular clusters \citep{Beasley2016a}. These early efforts led to estimates of a mass-to-light $(M/L)_g$ ratio of $\sim$ 50 --- 100 within the effective radius for DF44 in the Coma cluster and VCC 1287 in the Virgo Cluster, respectively. To estimate masses for significantly larger sample of UDGs, others apply scaling relations, either between the host galaxy's globular cluster populations and dynamical mass \citep{Blakeslee1997, Peng2008, Spitler2009, Harris2013, Harris2017}
or between the structural parameters and mass \citep{Zaritsky2017, Lee2020}. Although less reliable than the direct dynamical measurements, these studies agree that UDGs span a wide range of masses, with most UDGs likely to be comparable in mass to standard dwarf galaxies \citep[e.g.,][] {Amorisco2018} but also that there is a population extending to larger total masses  \citep[$>10^{11}$ M$_\odot$;][]{Zaritsky2017, Forbes2020, Lee2020}.

As mass measurements have improved, 
some of the early estimates for the most iconic UDGs have been revised downward, potentially undercutting both of the reasons for why UDGs may be of interest. For example, higher fidelity 
stellar kinematic measurements and globular cluster counts now result in estimates
of $(M/L)_I \sim 26$ $M_\odot/L_\odot$ within the effective radius for DF44 \citep{vanDokkum2019b} and a total mass in the neighborhood of 10$^{11}$ M$_\odot$ rather than the original estimates that placed it closer to 10$^{12}$ M$_\odot$ \citep{vanDokkum2019b, Saifollahi2020}.
Even so, DF44 and other similarly sized UDGs are extraordinary. Compare them with another galaxy of total mass $\sim$10$^{11}$ M$_\odot$, the Large Magellanic Cloud \citep{erkal}.
The mass-to-light ratio within the half-light radius of the Large Magellanic Cloud (LMC) is only 4.6 (calculated using $r_{1/2}$ from \citealt{Drlica-Wagner2020}, the enclosed i-band luminosity from \citealt{Eskew2011}, and the HI rotation curve of \citealt{Kim1998})
in comparison to $26^{+7}_{-6}$ for DF44 in solar units \citep{vanDokkum2019b}.

Our goals for this study are twofold. First, we search for more examples of physically large, and presumably massive, UDGs. Second, we extend the study of spectroscopically-confirmed UDGs to the field environment. With this work we aim to 
(1) establish the characteristics of physically large UDGs, and
(2) determine whether  such systems exist outside the cluster environment.

To study the physical characteristics of massive UDGs in a field environment, we need to conduct an extensive, spectroscopic follow-up survey of candidates. However, due to the low-surface brightnesses of even the largest UDGs, for which the spectrograph slit can  capture more light, obtaining spectroscopy is expensive. On 8- and 10-meter class telescopes, integration times range between 3600-5400 seconds for the largest UDGs \citep{Kadowaki17,vanDokkum2015b}. As such, most spectroscopic surveys have utilized multi-object spectroscopy, targeting regions of high UDGs surface density, galaxy clusters  \citep{Alabi18,Ruiz18,Chilingarian2019}. While this technique allows for  aggregations of spectra in a few observations, these surveys favor the ``typical'' UDG with smaller radius and thus smaller halo mass. Therefore, while there has been a significant increase in the number of spectroscopically-confirmed UDGs in the literature, confirmed field UDGs are still rare. We instead use single-object spectroscopy of UDG candidates that are potentially of the same large class as DF44 in the region surrounding the Coma cluster. We do this with the aim of investigating the properties of large UDGs over a contiguous region that spans environment classes. In \S\ref{sec:data} we present our observations and combine those with data available in the literature. In \S\ref{sec:discussion} we present our results, including the comparison of properties across environment. 


\section{Data}
\label{sec:data}

Our sample of low surface brightness candidate UDGs comes primarily from the first SMUDGes catalog \citep{Zaritsky19}, which was constructed using image processing techniques specifically designed to detect diffuse galaxies in survey images from the Legacy Survey \citep{dey}. That initial effort was carried out on the region encompassing the Coma galaxy cluster, where they could compare results to those from previous, independent UDG searches. To summarize, their technique employs a cleaning algorithm to eliminate image artifacts, point sources, and high surface brightness astronomical objects. After flattening these cleaned images, they apply wavelet filtering with varying kernel sizes to detect UDG candidates.  Selection on size and central surface brightness based on model fitting using GALFIT \citep{galfit1, galfit2} produces a candidate list that is further vetted through visual examination and a neural network classifier. Refinements to the image processing, parameter estimation, rejection of contamination, and completeness and uncertainty estimation are all in progress and may result in minor changes to the first SMUDGes catalog \citep{smudges2}. 

We augment the published SMUDGes catalog by similarly processing a few small regions of the Legacy Survey that lie far from the Coma cluster to provide targets when Coma was unavailable during our assigned telescope time. We refer to these areas and the candidates we identified there as ``off-Coma". These all have an RA $<$ 12$^h$. Spectroscopic observations of candidates in the off-Coma region yielded a few candidates with redshifts, but the bulk of our data come from the area covered by the original SMUDGes catalog, which we refer to as covering the ``Coma region".

We targeted candidates that are of large angular extent and lie beyond the Coma splashback radius. We impose the first of the two criteria because we seek physically large UDGs as indications suggest that they are hosted in more massive dark matter halos (see Appendix). Under the assumption that most of the candidates lie either within Coma or the surrounding large scale structure, those that have larger angular extent are also those that have larger physical extent. Of course, some contamination by physically smaller, nearby objects is expected. We impose the second of the two criteria because there is a growing sample of published redshifts for UDGs within the Coma cluster and because we seek to find field versions of the passive, large, massive UDG that are found within Coma. Due to practical concerns (right ascension distribution, weather), the observed sample is somewhat heterogeneous and in no way complete either in terms of angular size, environment, or surface brightness. 

We observed our selected sample of UDG candidates with the Multi-Object Double Spectrograph \citep[MODS;][]{Pogge2010} in binocular mode on the Large Binocular Telescope on three separate observing runs occurring on the nights of 2017 February 25-26, 2018 January 17-18, and 2019 April 4-5 (UTC). MODS1 experienced a glycol leak at the end of the 2017 February 25th (UTC) observing night, and was subsequently decommissioned for the remainder of the run. MODS2 remained functional throughout the 2017 February run.
Due to the flat surface brightness profiles of UDGs and their angular extent, we used a custom 2.4\arcsec-wide long slit, twice the width of the previously available widest slit, to increase the integrated light at the expense of spectral resolution. We positioned the long slit on our objects to maximize the integrated light from the galaxy, while primarily being constrained by the location of guide stars. When possible, we tried to minimize contamination from bright, nearby stars. We observed each target for 40 to 100 minutes in a series of dithered, 20- or 30-minute exposures (see Table \ref{table:observing_summary} for total exposure times), combining data from both spectrographs when possible. The total observing time for each object was dictated by both the surface brightness of the object and observing conditions. We used the G400L grating (400 lines mm$^{-1}$ blazed at $4.4^\circ$ centered on 4000 \AA\  with a resolution of 1850) in the blue channel (3200-5800 \AA) and the G670L grating (250 lines mm$^{-1}$ blazed at $4.3^\circ$ centered on 7600 \AA\  with a resolution of 2300) in the red channel (5800-10000 \AA). Because the red channel data suffer from higher sky backgrounds, we only use the blue channel data to measure redshifts.

In a series of steps, we prepared the data for analysis.
We used the {\tt modsCCDRed} package in the MODS data reduction pipeline \citep{Pogge2010} to fix bad pixels, subtract the bias level, correct for overscan and readout artifacts, and to flat field images. Next, we used the Laplacian Cosmic Ray Identification \citep[L.A. Cosmic;][]{vanDokkum2001} software suite to remove cosmic rays from the calibration and science images. We identified calibration lamp lines in our calibration spectra and calculated wavelength solutions for each observing run using the {\tt IDENTIFY}, {\tt REIDENTIFY}, and {\tt FITCOORDS} tasks in {\tt IRAF}\footnote{IRAF is distributed by the National Optical Astronomy Observatory, which is operated by the Association of Universities for Research in Astronomy (AURA) under a cooperative agreement with the National Science Foundation.} We refine our wavelength calibration using observed systematic zero-point deviations in prominent Hg(I) and [OI] night sky lines ($\lambda$=4046.56, 4358.35, 5460.74, and 5577.34 \AA) in MODS1 (2017) and MODS2 (2017 and 2018) blue-channel wavelength solutions. No refinement was required for MODS1 and MODS2 wavelength solutions for the 2019 data.
We correct for zero-point calibration offsets by applying the measured offset of 2.25 \AA\ and 1 \AA\ for the 2017 MODS1 and MODS2 blue-channel wavelength solution, respectively, and 0.53 \AA\  for MODS2 blue-channel in 2018. This correction sufficiently improves the precision of the wavelength solution to within 0.16 \AA. Even if one suspects, given the magnitude of the imposed zero-point velocity corrections, that the systematic wavelength calibration uncertainty is $\sim$ 2\AA, this level of error translates to a velocity uncertainty of $\sim$130 km s$^{-1}$, which does not affect any of our analysis. We applied the corrected wavelength solutions to our science images using the {\tt IRAF} task {\tt TRANSFORM}. We performed sky subtraction on all science frames using the {\tt BACKGROUND} task, with the lower ({\tt low\_rej}) and upper ({\tt high\_re}) rejection bounds set to 2-sigma and fitting {\tt order} interactively set to between 8 and 15. We then stacked all the exposures of each object using the {\tt IMSHIFT} and {\tt LSCOMBINE} {\tt IRAF} tasks. We could not always use the continuum from the UDG to align the individual exposures before stacking due to the low signal-to-noise of the spatially-extended low-surface-brightness profile. In those cases where the alignment was troublesome, we used stars that happened to lie in our 5$^\prime$ long slit. The long slit also enables us to use over 80\% of the slit length to obtain a high-S/N sky spectrum for subtraction. We performed a second sky subtraction pass once the images were stacked to remove any residual sky light or artifacts, using the same rejection parameters and a low fitting order ({\tt order}=3-4). Finally, we extracted the 1D stacked spectra for each object using the {\tt APALL IRAF} task. 

We repeatedly encountered scattered light in the MODS spectrograph that evolves on similar timescales as our exposure lengths. The light originates from bright ($m_r < 12$) stars withing 2 degrees of our targets. This scattered starlight dominates the continuum spectra of UDG candidates and sometimes happens to fall coincident on the detector to the UDG. To combat this problem, we eventually evolved our observing strategy to reject targets with scattered light present in the acquisition image taken immediately prior to the spectral observations. Furthermore, we excluded any individual  exposures with large areas of elevated counts, which we interpreted as a signature of scattered light, from the image stack.

We determined the redshift of our UDG candidates by cross-correlating the spectra of our targets with an A-type stellar template. We choose this particular template because of the prominence of Balmer lines in UDG spectra \citep{Kadowaki17}. We also tested a later type stellar template, but in no case where the A-type template failed did the later-type template result in an identified redshift. In the cross-correlation, we exclude a list of narrow wavelength ranges that correspond to bright sky lines and laser light contamination in MODS (the latter is used to enable flexure compensation while observing). We measured the redshift of each object using the {\tt XCSAO} and the {\tt EMSAO IRAF} tasks in the {\tt RVSAO} add-on package, setting the acceptable recessional velocity range between $-$5000 to 30000 km s$^{-1}$. We visually examined each fit. Of the targets listed in Table \ref{table:observing_summary}, only 14 out of 23 yielded acceptable redshifts. UDG redshift estimation with these low S/N data is somewhat uncertain and catastrophic failures are possible. While we only present those in which we have confidence, it is certainly possible that in a small number of cases we have been fooled by an unlucky coincidence of noise.

While the continuum was present in the 2D spectra of all 23 observed objects, most of those that did not yield redshifts were dominated by noise despite the long exposure times. In a handful of cases, we observed no discernible emission or absorption lines despite a reasonably strong continuum. In the remainder of cases, despite our efforts we were foiled by low-levels of scattered light.


\begin{table*}
\begin{center}
\caption{LBT Observation Log
\label{table:observing_summary}}
\begin{tabular}{lrlrrll}
\tableline
\\
    UDG
    & t$_{exp}$
    & Date
    & Seeing
    & Airmass
    & Cloud 
    & Notes
    \\
    & (min)
    & (UTC)
    & ('')
    &
    & Coverage
\\
\tableline 
\\
SMDG0239472+011236   & 120 & 2018 Jan 18 & 1.2 & 1.25 & Fairly clear & Faint, Scattered light \\
SMDG0244338$-$001602 &  80 & 2018 Jan 17 & 0.8 & 1.48 & Thin clouds  & \\
                     & 120 & 2018 Jan 18 & 1.5 & 1.20 & Thin clouds  & \\
SMDG0838589+260428   & 120 & 2017 Feb 25 & 1.9 & 1.13 & Fairly clear & Faint \\
SMDG0852477+324943   & 115 & 2018 Jan 17 & 0.8 & 1.03 & Thin clouds  & Faint \\
SMDG0854195+310242   & 151 & 2018 Jan 18 & 0.9 & 1.34 & Fairly clear & Faint, Scattered light \\
SMDG0855549+312822   & 134 & 2018 Jan 18 & 0.8 & 1.02 & Fairly clear & \\ 
SMDG0856259+315502   & 119 & 2018 Jan 17 & 0.6 & 1.13 & Thin clouds  & \\
SMDG0914401+283036   &  40 & 2017 Feb 26 & 1.5 & 1.02 & Fairly clear & \\
SMDG1006234+285218   & 120 & 2017 Feb 25 & 1.5 & 1.06 & Fairly clear & \\
SMDG1216089+325257   & 180 & 2019 Apr 05 & 0.9 & 1.42 & Clear & Scattered light \\ 
SMDG1217377+283519   & 120 & 2017 Feb 25 & 1.0 & 1.22 & Fairly clear & \\
                     &  60 & 2017 Feb 26 & 1.7 & 1.17 & Fairly clear & \\
SMDG1218390+285050   & 200 & 2018 Jan 17 & 0.9 & 1.20 & Clear        & \\
SMDG1221086+292921   & 120 & 2017 Feb 25 & 1.3 & 1.04 & Fairly clear & Faint \\ 
SMDG1224081+280545   & 120 & 2017 Feb 25 & 1.2 & 1.15 & Fairly clear & Scattered light \\ 
SMDG1237294+204442   & 120 & 2018 Jan 17 & 0.8 & 1.45 & Thin clouds  & \\
SMDG1238305+274355   &  40 & 2019 Apr 04 & 1.0 & 1.57 & Fairly clear & Scattered light \\
                     &  90 & 2019 Apr 05 & 1.1 & 1.05 & Clear        & Scattered light \\
SMDG1240530+321655   & 180 & 2019 Apr 04 & 1.1 & 1.31 & Fairly clear & \\
SMDG1242314+315809   & 180 & 2019 Apr 04 & 1.8 & 1.17 & Thin clouds  & Scattered light  \\ 
SMDG1245277+181801   & 120 & 2018 Jan 17 & 0.5 & 1.09 & Thin clouds  & \\
                     & 120 & 2019 Apr 05 & 1.1 & 1.13 & Clear        & Scattered light  \\ 
SMDG1247231+180142   & 180 & 2019 Apr 04 & 1.5 & 1.07 & Thin clouds  & \\
SMDG1251014+274753   & 120 & 2017 Feb 25 & 1.4 & 1.03 & Fairly clear & \\ 
SMDG1304536+274252   &  40 & 2017 Feb 24 & 2.0 & 1.10 & Fairly clear & \\
SMDG1335454+281225   &  80 & 2018 Jan 17 & 0.8 & 1.02 & Clear        &  
\\
\tableline
\end{tabular}
\end{center}
\end{table*}

We present the parameters of our sample galaxies, including the physical parameters that can be calculated once distances are known, and details of the observations in a series of tables.
To calculate physical quantities, we adopt a $\Lambda$CDM cosmology with $\Omega_\textrm{M}$=0.31, $\Omega_\Lambda$=0.69, and $H_0=67 \, \textrm{km s}^{-1} \textrm{Mpc}^{-1}$ \citep{planck2018}.
In Tables \ref{table:LBT_targets}, \ref{table:Non_LBT_targets1}, and \ref{table:Non_LBT_targets2}, we provide the $g$-band central surface brightness ($\mu_g(0)$), effective radius ($r_\mathrm{e}$), $g$-band absolute magnitude ($M_g$), recessional velocity ($cz$), and environmental classifications (discussed below) for the set of UDG candidates in the Coma region and in our off-Coma region.
The tables are divided according to the provenance of both the spectroscopic redshift and the candidates. In Table \ref{table:LBT_targets} we present data for  SMUDGes candidates for which we obtained redshifts with the LBT, as described here and by \cite{Kadowaki17}.
In Table \ref{table:Non_LBT_targets1} we present the data for SMUDGes candidates in the Coma region for which redshifts come from other sources. Finally, in Table \ref{table:Non_LBT_targets2} we present the data for UDG candidates in the Coma region that are not in the SMUDGes catalog and for which the redshifts come from other sources.

We report all redshifts in Tables \ref{table:LBT_targets}, \ref{table:Non_LBT_targets1}, and \ref{table:Non_LBT_targets2} in the CMB rest frame. Specifically, we used the {\tt radial\_velocity\_correction} method in {\tt Astropy}'s {\tt SkyCoord} module to correct for heliocentric velocities of objects in Table \ref{table:observing_summary} and \cite{Kadowaki17}. While \cite{vanDokkum2015b}, \cite{Gu2018}, \cite{Ruiz18}, and \cite{Chilingarian2019} do not explicitly state the adopted frame, we assume that the reported velocities are heliocentric on the basis of some inter-comparison among published results. Velocities from all other literature sources explicitly state that they are corrected to the heliocentric reference frame.
We adopt apex of $\ell_\text{apex}=264.14^\circ$, $b_\text{apex}=+48.26^\circ$, and $V_\text{apex}=371.0 \, \mathrm{km \, s}^{-1}$ \citep{Fixsen1996}
to compute the CMB rest frame velocities for all galaxies. 
Additionally, we use only the whole galaxy ``W'' velocity measurements from \cite{Ruiz18}. In the cases where multiple literature sources report velocity measurements for the same galaxy, we first compute the heliocentric, weighted mean velocity and its standard error with variance weighting before converting the velocity into the CMB rest frame.

We report the redshift uncertainties when possible. We note that the spectroscopic redshift uncertainties for some literature sources were unavailable. When available, we propagate the redshift uncertainties through our determination of the overall uncertainties for $r_e$, $M_\text{NUV}$, $M_g$, $M_r$, and $M_z$, with optical uncertainties provided by GALFIT. The average uncertainty in $r_e$ is $\sim$ 0.1 kpc, although it trends larger at higher redshifts. Six candidates at roughly the distance of Coma have uncertainties greater than 0.25 kpc, while two candidates in the background to Coma have the largest $r_e$ uncertainties of 0.5 and 1 kpc. Uncertainties in the magnitudes are from 0.05 to 0.1 for each band, with the uncertainty decreasing at higher redshifts because of the lower relative uncertainty in the distance. All 6 objects with magnitude uncertainties greater than 0.2 occur in candidates with recessional velocities less than 1600 km s$^{-1}$. Uncertainties at these levels do not affect the broad conclusions we discuss below. Instead, uncertainties are dominated by sample size, statistical considerations, and selection biases.

A few details are important. 
We report the extinction-corrected photometry in Tables \ref{table:LBT_targets}, \ref{table:Non_LBT_targets1}, and \ref{table:Non_LBT_targets2}. We measured the structural parameters and the optical, grz-band photometry for all sources, even those not originally in the SMUDGes catalog, using the standard SMUDGes pipeline described by \cite{Zaritsky19}. One major change to the photometry pipeline allowed for floating S\'ersic indices $n$. Compared to the previous version, which strictly fixed $n$ to 1 for exponential profiles, the floating $n$ pipeline yields more accurate photometry.
We used NED's Galactic Extinction Calculator\footnote{ned.ipac.caltech.edu/forms/calculator.html} to correct for extinction, using the extinction map from \cite{Schafly2011}. As such, all of the candidates are on the same photometric system and analyzed in the same manner.

Similarly, we redo the NUV photometry from GALEX \citep{martin2005} data obtained via MAST\footnote{NASA's public archive portal accessible at archive.stsci.edu} both because some of the candidates we consider here are not in the original SMUDGes catalog and because our new fitting has changed the measured $r_e$ in those what were included in the original work. We follow the measurement procedure presented by \cite{rs} and consider only those with flux signal-to-noise (S/N) estimates $>$ 3 to be detected.
Among the 68 candidates, four  (SMDG1237294+204442, SMDG1306148+275941, SMDG1313189+312452, and SMDG1315427+311846) have previously reported NUV detections \citep{rs} and we recover three of those as well. We do not recover SMDG1306148+275941, which also happens to be the only object in our catalog with a prior FUV detection, but it lies only slightly below our S/N $>$ 3 criterion. Two of the candidates lack {\sl GALEX} data (SMDG1231329+232917 and SMDG1253152+274115). 
In total, we detect 20 of our candidate UDGs in the NUV band. 10 of these candidate UDGs (SMDG1313189+312452, SMDG1221577+281436, SMDG1226040+241802, SMDG1230359+273310, SMDG1241425+273352, SMDG1248019+261235, SMDG1255415+191239, SMDG1302418+215952, SMDG1312223+312320, and SMDG1315427+311846) are detected in 100-second AIS-images; 6 (SMDG1103517+284118, SMDG1240017+261920, SMDG1239050+323015, SMDG1300263+272735, SMDG1220188+280132, and SMDG1225185+270858) are detected in guest investigator programs; and an additional 4 (SMDG0914401+283036, SMDG1237294+204442, SMDG1245277+181801, and SMDG1225277+282903) are detected in 1500-second MIS-images. We present the NUV magnitudes or our upper limits in Tables \ref{table:LBT_targets} to \ref{table:Non_LBT_targets2}.

\begin{figure*}
\includegraphics[width=0.95\textwidth]{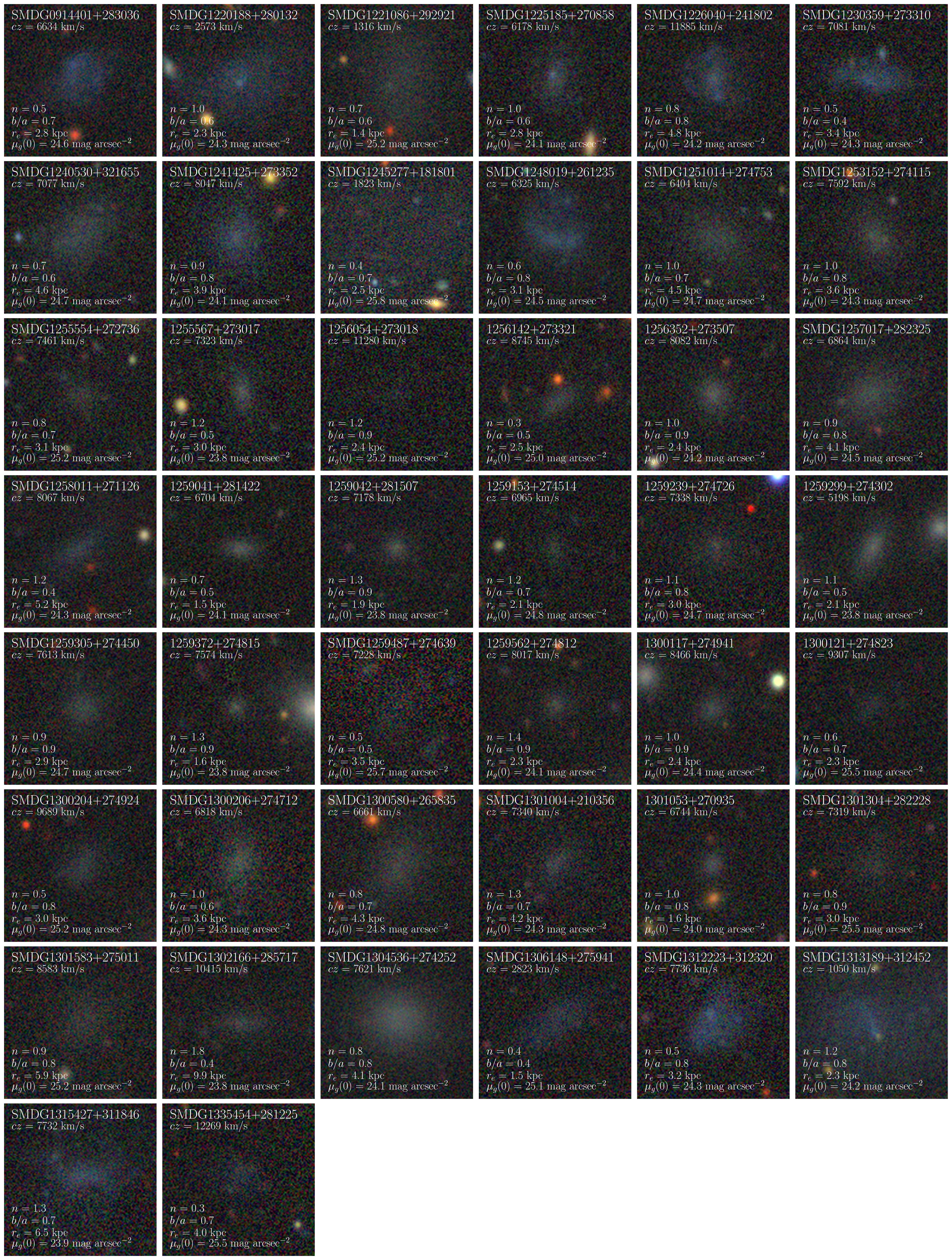}
\caption{Legacy Survey images of the 44 spectroscopically-confirmed UDGs in our sample. All but one (SMDG0914401+283036)
are projected within $11^\circ$ of the center of the Coma Cluster. We include the recessional velocity $cz$, S\'{e}rsic index $n$, axis ratio $b/a$, effective radius $r_e$, and g-band, central surface brightness $\mu_g(0)$ for each object. Images are each 32\arcsec $\times$ 32\arcsec, with North at the top and East to the left.}
\label{fig:udgs}
\end{figure*}

In the subset of candidates with LBT spectroscopy, we observe bright [O II] and [O III] emission lines only in the spectra of SMDG1301304+282228 (DF08) and the off-Coma UDG SMDG0914401+283036. In the latter, the optical flux likely  originates from two distinct regions of the galaxy that can be seen as bright patches in Figure \ref{fig:udgs}. We also report a NUV flux detection for this source, consistent with the interpretation of the emission line flux as indicative of ongoing star formation. SMDG1301304+282228 (DF08) is a bit more puzzling in that we have only a NUV flux limit and it falls on the optical red sequence (Figure 2 from \citealp{rs}), both of which would suggest that it is passive. DF08 serves as a cautionary tale that no single star formation indicator can definitively identify objects with low levels of active or recent star formation. Low levels of star formation may very well be occurring even in objects with NUV limits and red optical colors.


\begin{rotatetable*}
\begin{deluxetable*}{lccrrrrcccrrr}
\movetableright=0.1cm
\tablecolumns{13}
\tablecaption{LBT SMUDGes Redshift Sample}
\label{table:LBT_targets}
\tablehead{
    \colhead{Name}
    & \colhead{Alternate\note{1}}
    & \colhead{$\mu_g(0)$}
    & \colhead{$M_\mathrm{NUV}$}
    & \colhead{$M_g$}
    & \colhead{$M_r$}
    & \colhead{$M_z$}
    & \colhead{$r_\mathrm{e}$}
    & \colhead{$b/a$}
    & \colhead{$n$}
    & \colhead{$cz_{\scaleto{\text{CMB}}{2.5pt}}$\note{2}}
    & \colhead{Local Env.\note{3}}
    & \colhead{Cluster?}
    \\
    & \colhead{Name}
    & \colhead{(mag $\square\arcsec$)}
    & \colhead{(mag)}
    & \colhead{(mag)}
    & \colhead{(mag)}
    & \colhead{(mag)}
    & \colhead{(kpc)}
    & 
    & 
    & \colhead{(km s$^{-1}$)}
    & 
    &
    \\
}
\startdata 
SMDG0244338$-$001602         &               & 25.4 & $>$- 9.8 & -12.4 & -12.9 & -13.3 & 0.9 & 0.86 & 0.56 & $ 1282 \pm  52$ & D  & No  \\
SMDG0855549+312822\note{4}   &               & 24.6 & $>$- 7.7 &  -9.7 & -10.2 & -10.6 & 0.3 & 0.55 & 0.92 & $  771 \pm  87$ & U  & No  \\
SMDG0914401+283036           &               & 24.6 &    -14.5 & -15.6 & -15.8 & -15.9 & 2.8 & 0.71 & 0.46 & $ 6634 \pm  51$ & S  & No  \\
SMDG1006234+285218           &               & 25.1 & $>$- 9.7 & -12.0 & -12.5 & -12.8 & 0.8 & 0.69 & 0.67 & $ 1594 \pm  51$ & S  & No  \\
SMDG1216089+325257           &               & 25.8 & $>$- 6.0 & -10.1 & -10.6 & -11.1 & 0.4 & 0.89 & 0.60 & $  864 \pm 100$ & U  & No  \\
SMDG1217377+283519           &               & 24.9 & $>$- 8.4 & -11.6 & -12.1 & -12.4 & 0.6 & 0.76 & 0.80 & $  790 \pm  69$ & U  & No  \\
SMDG1221086+292921           &               & 25.2 & $>$-10.2 & -12.9 & -13.5 & -13.8 & 1.4 & 0.59 & 0.72 & $ 1316 \pm  66$ & D  & No  \\
SMDG1237294+204442           &               & 24.8 &    -10.4 & -12.0 & -12.2 & -12.3 & 0.6 & 0.64 & 0.55 & $ 1751 \pm  48$ & S  & No  \\
SMDG1240530+321655           &               & 24.7 & $>$-12.0 & -16.1 & -16.6 & -16.9 & 4.6 & 0.62 & 0.70 & $ 7077 \pm  38$ & S  & No  \\
SMDG1242314+315809           &               & 24.8 & $>$- 7.6 & -11.1 & -11.7 & -12.0 & 0.4 & 0.76 & 0.66 & $  826 \pm  63$ & U  & No  \\
SMDG1245277+181801\note{4}   &               & 25.8 &    -12.5 & -14.2 & -14.5 & -14.7 & 2.5 & 0.72 & 0.40 & $ 1823 \pm 114$ & S  & No  \\
SMDG1251014+274753           &               & 24.7 & $>$-12.7 & -15.8 & -16.4 & -16.7 & 4.5 & 0.73 & 0.97 & $ 6404 \pm  45$ & S  & No  \\
SMDG1253152+274115\note{4}   &          DF30 & 24.3 &          & -15.6 & -16.1 & -16.6 & 3.6 & 0.76 & 1.00 & $ 7592 \pm  81$ & D  & No  \\
SMDG1257017+282325\note{5,6} & Yagi680, DF07 & 24.5 & $>$-11.5 & -16.0 & -16.6 & -17.0 & 4.1 & 0.76 & 0.86 & $ 6864 \pm  33$ & D  & Yes \\
SMDG1258011+271126\note{5}   & Yagi507, DF40 & 24.3 & $>$-13.1 & -15.5 & -15.9 & -16.2 & 5.2 & 0.44 & 1.18 & $ 8067 \pm  46$ & D  & Yes \\
SMDG1301304+282228\note{5}   & Yagi194, DF08 & 25.5 & $>$-10.9 & -14.6 & -15.3 & -15.5 & 3.0 & 0.93 & 0.77 & $ 7319 \pm  97$ & D  & Yes \\
SMDG1302166+285717\note{5}   & Yagi215, DF03 & 23.8 & $>$-12.9 & -16.2 & -16.7 & -17.1 & 9.9 & 0.37 & 1.79 & $10415 \pm  37$ & D  & No  \\
SMDG1304536+274252           &               & 24.1 & $>$-12.6 & -16.7 & -17.3 & -17.6 & 4.1 & 0.84 & 0.77 & $ 7621 \pm  61$ & D  & Yes \\
SMDG1335454+281225           &               & 25.5 & $>$-12.8 & -15.6 & -15.8 & -15.9 & 4.0 & 0.69 & 0.33 & $12269 \pm 190$ & S  & No  \\
\\
\enddata
\tablenotetext{1}{Alternate names sourced from \cite{vanDokkum2015a} and \cite{Yagi2016}. Yagi objects are also denoted as ``Subaru-UDG'' in NASA's Extragalactic Database.}
\tablenotetext{2}{Redshifts converted to CMB rest frame.}
\tablenotetext{3}{The local environment designations are D for dense, S for sparse, and U for unconstrained.}
\tablenotetext{4}{Redshift identification is somewhat uncertain.}
\tablenotetext{5}{Observer-frame redshift was previously reported in \cite{Kadowaki17}.}
\tablenotetext{6}{Heliocentric redshift was previously reported in \cite{Gu2018}. We report the CMB rest frame-corrected, variance-weighted, mean velocity and standard error taken from measurements for our LBT measurements and those reported in \cite{Gu2018}.}
\end{deluxetable*}
\end{rotatetable*}


\begin{longrotatetable}
\begin{deluxetable*}{lccrrrrcccrrrr}
\movetableright=0.1cm
\tablecolumns{14}
\tablewidth{0pt}
\tablecaption{Non-LBT SMUDGes Redshift Sample}
\label{table:Non_LBT_targets1}
\tablehead{
    \colhead{Name}
    & \colhead{Alternate\note{1}}
    & \colhead{$\mu_g(0)$}
    & \colhead{$M_\mathrm{NUV}$}
    & \colhead{$M_g$}
    & \colhead{$M_r$}
    & \colhead{$M_z$}
    & \colhead{$r_\mathrm{e}$}
    & \colhead{$b/a$}
    & \colhead{$n$}
    & \colhead{$cz_{\scaleto{\text{CMB}}{2.5pt}}$\note{2}}
    & \colhead{Local Env.\note{3}}
    & \colhead{Cluster?}
    & \colhead{Ref.\note{4}}
    \\
    & \colhead{Name}
    & \colhead{(mag $\square\arcsec$)}
    & \colhead{(mag)}
    & \colhead{(mag)}
    & \colhead{(mag)}
    & \colhead{(mag)}
    & \colhead{(kpc)}
    & 
    & 
    & \colhead{(km s$^{-1}$)}
    & 
    & 
    &
    \\
}
\startdata
SMDG1103517+284118 &                & 25.7 &    -10.6 & -12.4 & -12.6 & -12.7 & 1.1 & 0.89 & 0.69 & $  969 \pm   2$ & U & No  & 2  \\
SMDG1220188+280132 &                & 24.3 &    -13.1 & -14.6 & -14.9 & -15.0 & 2.3 & 0.65 & 0.98 & $ 2573 \pm   2$ & S & No  & 2  \\
SMDG1221577+281436 &                & 25.1 &     -9.5 & -12.2 & -12.8 & -13.2 & 0.9 & 0.58 & 0.60 & $  738 \pm   8$ & U & No  & 1  \\
SMDG1223451+283550 &                & 24.2 &  $>$-9.0 & -12.9 & -13.4 & -13.7 & 0.9 & 0.60 & 0.72 & $ 2663 \pm   4$ & S & No  & 2  \\
SMDG1225185+270858 &                & 24.1 &    -13.4 & -15.0 & -15.5 & -15.8 & 2.8 & 0.63 & 1.04 & $ 6178 \pm   5$ & S & No  & 2  \\
SMDG1225277+282903 &                & 24.0 &    -12.9 & -13.5 & -13.5 & -13.5 & 0.8 & 0.82 & 0.57 & $  769 \pm 0.2$ & U & No  & 3  \\
SMDG1226040+241802 &                & 24.2 &    -14.4 & -16.7 & -17.1 & -17.3 & 4.8 & 0.81 & 0.85 & $11885 \pm   3$ & S & No  & 2  \\
SMDG1230359+273310 &                & 24.3 &    -14.0 & -15.6 & -15.9 & -16.0 & 3.4 & 0.40 & 0.49 & $ 7081 \pm   3$ & S & No  & 2  \\
SMDG1231329+232917 &                & 24.0 &          & -13.0 & -13.2 & -13.2 & 1.0 & 0.48 & 0.88 & $ 1360 \pm   6$ & S & No  & 2  \\
SMDG1239050+323015 &                & 24.0 &    -10.3 & -11.7 & -11.8 & -11.9 & 0.6 & 0.85 & 1.36 & $  877 \pm   1$ & U & No  & 2  \\
SMDG1240017+261920 &                & 24.5 &    -10.8 & -11.8 & -11.9 & -11.9 & 0.5 & 0.69 & 0.54 & $  738 \pm   2$ & U & No  & 2  \\
SMDG1241425+273352 &                & 24.1 &    -14.4 & -16.3 & -16.7 & -16.9 & 3.9 & 0.78 & 0.85 & $ 8047 \pm   2$ & S & No  & 2  \\
SMDG1248019+261235 &                & 24.5 &    -13.9 & -15.8 & -16.1 & -16.2 & 3.1 & 0.79 & 0.61 & $ 6325 \pm   5$ & D & No  & 2  \\
SMDG1253571+291500 &                & 24.5 & $>$- 7.1 & -10.5 & -11.3 & -11.5 & 0.4 & 0.52 & 0.69 & $  770 \pm   4$ & U & No  & 2  \\
SMDG1255415+191239 &                & 24.2 &    -12.4 & -14.4 & -14.5 & -14.3 & 2.2 & 0.45 & 0.79 & $  718 \pm   1$ & U & No  & 2  \\
SMDG1255554+272736 & Yagi762, DF36  & 25.2 & $>$-11.7 & -14.6 & -15.3 & -15.7 & 3.1 & 0.72 & 0.85 & $ 7461 \pm 127$ & D & Yes & 8  \\
SMDG1259305+274450 & Yagi276, DF28  & 24.7 & $>$-10.9 & -15.1 & -15.8 & -16.1 & 2.9 & 0.86 & 0.88 & $ 7613 \pm 102$ & D & Yes & 8  \\
SMDG1259487+274639 & Yagi285, DF25  & 25.7 & $>$-11.7 & -14.5 & -14.8 & -14.8 & 3.5 & 0.53 & 0.54 & $ 7228 \pm 121$ & D & Yes & 8  \\
SMDG1300204+274924 & Yagi090        & 25.2 & $>$-12.1 & -15.2 & -15.7 & -16.0 & 3.0 & 0.76 & 0.53 & $ 9689 \pm  42$ & D & Yes & 9  \\
SMDG1300206+274712 & Yagi093, DF26  & 24.3 & $>$-12.5 & -15.6 & -16.3 & -16.6 & 3.7 & 0.65 & 0.95 & $ 6818 \pm  27$ & D & Yes & 9  \\
SMDG1300263+272735 &                & 23.0 &    -13.7 & -16.9 & -17.2 & -17.5 & 3.7 & 0.44 & 0.75 & $ 7209 \pm   2$ & D & Yes & 7  \\
SMDG1300580+265835 & Yagi011, DF44  & 24.8 & $>$- 8.9 & -15.7 & -16.3 & -16.7 & 4.3 & 0.68 & 0.83 & $ 6661 \pm  38$ & S & Yes & 5, 6  \\
SMDG1301004+210356 &                & 24.3 & $>$-12.7 & -15.6 & -16.0 & -16.2 & 4.2 & 0.68 & 1.30 & $ 7340 \pm   3$ & S & No  & 2  \\
SMDG1301583+275011 & Yagi165, DF17  & 25.2 & $>$-13.6 & -16.0 & -16.6 & -16.9 & 5.9 & 0.77 & 0.93 & $ 8583 \pm  43$ & S & Yes & 6  \\
SMDG1302418+215952 &                & 24.0 &    -10.0 & -13.2 & -13.6 & -13.9 & 0.7 & 0.86 & 0.64 & $  553        $ & U & No  & 4  \\
SMDG1306148+275941 &                & 25.1 & $>$-11.2 & -13.2 & -13.5 & -13.6 & 1.5 & 0.45 & 0.44 & $ 2823 \pm   3$ & D & No  & 2  \\
SMDG1312223+312320 &                & 24.3 &    -14.8 & -16.2 & -16.5 & -16.5 & 3.2 & 0.75 & 0.49 & $ 7736 \pm   3$ & S & No  & 2  \\
SMDG1313189+312452 &                & 24.2 &    -10.8 & -14.4 & -14.7 & -14.9 & 2.3 & 0.78 & 1.23 & $ 1050 \pm   5$ & U & No  & 2  \\
SMDG1315427+311846 &                & 23.9 &    -15.5 & -16.7 & -17.1 & -17.2 & 6.5 & 0.70 & 1.30 & $ 7732 \pm   6$ & S & No  & 2  \\
\\
\enddata
\tablenotetext{1}{Alternate names sourced from \cite{vanDokkum2015a} and \cite{Yagi2016}. Yagi objects are also denoted as ``Subaru-UDG'' in NASA's Extragalactic Database.}
\tablenotetext{2}{Redshifts converted to CMB rest frame.}
\tablenotetext{3}{The local environment designations are D for dense, S for sparse, and U for unconstrained.}
\tablenotetext{4}{References used to source redshifts only and} are designated as follows: (1) \cite{Huchtmeier03}; (2) \cite{karunakaran20}; (3) \cite{Huchtmeier09}; (4) \cite{Kim14}; (5) \cite{vanDokkum2015b}; (6) \cite{Gu2018}; (7) \cite{Chilingarian2019}; (8) \cite{Alabi18}; (9) \cite{Ruiz18}
\end{deluxetable*}
\end{longrotatetable}


\begin{rotatetable*}
\begin{deluxetable*}{lccrrrrcccrrrr}
\movetableright=0.05cm
\tablewidth{0pt}
\tablecaption{Non-LBT non-SMUDGes Redshift Sample}
\label{table:Non_LBT_targets2}
\tablehead{
    \colhead{Name}
    & \colhead{Alternate\note{1}}
    & \colhead{$\mu_g(0)$}
    & \colhead{$M_\mathrm{NUV}$}
    & \colhead{$M_g$}
    & \colhead{$M_r$}
    & \colhead{$M_z$}
    & \colhead{$r_\mathrm{e}$}
    & \colhead{$b/a$}
    & \colhead{$n$}
    & \colhead{$cz_{\scaleto{\text{CMB}}{2.5pt}}$\note{2}}
    & \colhead{Local Env.\note{3}}
    & \colhead{Cluster?}
    & \colhead{Ref.\note{4}}
    \\
    & \colhead{Name}
    & \colhead{(mag $\square\arcsec$)}
    & \colhead{(mag)}
    & \colhead{(mag)}
    & \colhead{(mag)}
    & \colhead{(mag)}
    & \colhead{(kpc)}
    & 
    & 
    & \colhead{(km s$^{-1}$)}
    & 
    & 
    &
    \\
}
\startdata
1255567+273017         & Yagi764        & 23.8 & $>$-12.0 & -14.9 & -15.6 & -15.9 & 3.0 & 0.48 & 1.23 & $ 7323 \pm 115$ & D  & Yes & 8  \\
1256054+273018         & Yagi771        & 25.2 & $>$-12.8 & -13.8 & -14.2 & -14.6 & 2.4 & 0.91 & 1.18 & $11280 \pm 192$ & S  & No  & 8  \\
1256142+273321\note{5} & Yagi776        & 25.0 & $>$-11.3 & -14.8 & -15.4 & -16.0 & 2.5 & 0.54 & 0.34 & $ 8745 \pm  81$ & S  & Yes & 8  \\
1256352+273507         & Yagi786        & 24.2 & $>$-11.4 & -15.2 & -16.0 & -16.4 & 2.4 & 0.90 & 0.95 & $ 8082 \pm 141$ & S  & Yes & 8  \\
1259041+281422         & Yagi452        & 24.1 &  $>$-9.5 & -14.1 & -14.7 & -15.1 & 1.5 & 0.54 & 0.70 & $ 6704 \pm   5$ & D  & Yes & 7  \\
1259042+281507         & Yagi343        & 23.8 &  $>$-9.6 & -14.4 & -15.1 & -15.5 & 1.9 & 0.93 & 1.32 & $ 7178 \pm   6$ & D  & Yes & 7  \\
1259153+274514         & Yagi263        & 24.8 & $>$-10.3 & -13.5 & -14.3 & -14.4 & 2.1 & 0.66 & 1.16 & $ 6965 \pm 147$ & D  & Yes & 8  \\
1259239+274726         & Yagi364, DF23  & 24.7 & $>$-12.7 & -14.8 & -15.4 & -15.8 & 3.0 & 0.84 & 1.15 & $ 7338 \pm  90$ & D  & Yes & 8  \\
1259299+274302         & Yagi275        & 23.8 & $>$-10.6 & -14.6 & -15.3 & -15.6 & 2.1 & 0.54 & 1.06 & $ 5198 \pm   4$ & D  & Yes & 7, 8  \\
1259372+274815         & Yagi376        & 23.8 & $>$-10.3 & -14.1 & -14.7 & -15.1 & 1.6 & 0.93 & 1.25 & $ 7574 \pm   5$ & D  & Yes & 7  \\
1259562+274812         & Yagi392        & 24.1 & $>$-10.3 & -14.5 & -15.1 & -15.6 & 2.3 & 0.92 & 1.36 & $ 8017 \pm 161$ & D  & Yes & 8  \\
1300004+274819         & Yagi398        & 23.6 &  $>$-8.8 & -13.7 & -14.4 & -14.8 & 1.1 & 0.97 & 1.19 & $ 4449 \pm 167$ & S  & No  & 8  \\
1300054+275333         & Yagi407        & 23.4 & $>$-10.7 & -15.0 & -15.7 & -16.1 & 2.3 & 0.90 & 1.39 & $ 6553 \pm   4$ & D  & Yes & 7  \\
1300117+274941         & Yagi418        & 24.4 & $>$-10.9 & -14.8 & -15.4 & -15.9 & 2.4 & 0.87 & 0.99 & $ 8466 \pm  40$ & D  & Yes & 8, 9  \\
1300121+274823         & Yagi417        & 25.5 & $>$-10.4 & -14.1 & -14.5 & -15.0 & 2.3 & 0.72 & 0.60 & $ 9307 \pm 179$ & S  & Yes & 8  \\
1300232+275225\note{5} &                & 22.7 & $>$-11.6 & -15.4 & -16.0 & -16.4 & 1.5 & 0.85 & 1.07 & $ 7249 \pm  19$ & D  & Yes & 9  \\
1300243+275155         &                & 23.3 & $>$-11.5 & -14.9 & -15.6 & -16.0 & 1.9 & 0.83 & 1.24 & $ 6636 \pm  19$ & D  & Yes & 9  \\
1300284+274820         & Yagi106        & 23.4 & $>$-11.5 & -14.8 & -15.4 & -15.8 & 2.2 & 0.66 & 1.29 & $ 6740 \pm   4$ & D  & Yes & 7  \\
1300387+272835\note{6} &                & 22.5 & $>$-12.5 & -16.2 & -16.7 & -17.0 & 2.4 & 0.56 & 1.06 & $ 8207 \pm   3$ & D  & Yes & 7  \\
1301053+270935         & Yagi012, DFX2  & 24.0 & $>$-10.7 & -14.1 & -14.6 & -15.1 & 1.6 & 0.78 & 1.02 & $ 6744 \pm  33$ & S  & Yes & 8  \\
\\
\enddata
\tablenotetext{1}{Alternate names sourced from \cite{vanDokkum2015a} and \cite{Yagi2016}, with the exception of DFX2 which is sourced from \cite{Alabi18}. Yagi objects are also denoted as ``Subaru-UDG'' in NASA's Extragalactic Database.}
\tablenotetext{2}{Redshifts converted to CMB rest frame.}
\tablenotetext{3}{The local environment designations are D for dense, S for sparse, and U for unconstrained.}
\tablenotetext{4}{References used to source redshifts only and are designated as in Table \ref{table:Non_LBT_targets1}.}
\tablenotetext{5}{This object is too faint to be modeled in the Legacy Survey data.}
\tablenotetext{6}{Galaxy is brighter than our pipeline threshold and was masked during processing.}
\end{deluxetable*}
\end{rotatetable*}

\twocolumngrid


\section{Results}
\label{sec:discussion}

Before proceeding to present our results, we discuss one subtlety in our UDG criteria. Given observational uncertainties in $\mu_g(0)$ and $r_e$, it was unclear whether we should exclude objects that were within their measurement uncertainties of our class-defining thresholds of $\mu_g(0) \ge 24$ mag arcsec$^{-2}$ and $r_e \geq 1.5$ kpc. We opted to 
 also consider as UDGs those galaxies that do not strictly satisfy the criteria but are within 2$\sigma$ of the surface brightness and size thresholds. These modified criteria can also be thought of as effectively lowering the size and surface brightness cuts for our sample to $r_e \geq 1.4$ kpc and $\mu_g(0) \geq 23.8$ mag arcsec$^{-2}$. This choice leads to an overall modest increase of 9 additional UDGs to our sample, up to a total of 44 confirmed UDGs (SMDG1221086+292921, SMDG1302166+285717, SMDG1306148+275941, SMDG1315427+311846, 1255567+273017, 1259041+281422,  1259042+281507, 1259299+274302, 1259372+274815). Although we have modestly relaxed the defining criteria of a UDG, we do not account for measurement uncertainties when we refer to \textit{large} ($\geq3.5$ kpc) UDGs.

Once distances are established, many of the UDG candidates fail the physical size criterion for the UDG class. Subsequently, we refer to candidates that pass the size criterion as UDGs and to those that fail as non-UDGs. Given that the physically largest UDGs are rare, a spectroscopic selection favoring candidates of large angular size, which is the case in our work, will naturally favor nearer, smaller galaxies. Of the 19 candidates for which we present LBT redshifts in Table \ref{table:LBT_targets}, four have $cz_\text{CMB} < 1000$ km s$^{-1}$ and an additional 5 have $cz_\text{CMB} < 2000$ km s$^{-1}$. Among all the candidates we present in the three tables, 12 have $cz_\text{CMB} < 1000$ km s$^{-1}$. Due to peculiar velocities, we cannot reliably compute the distances of UDGs candidates with small recessional velocities.
We will not consider further the characteristics of galaxies with $cz_\text{CMB} < 1000$ km s$^{-1}$, except to say that all but one (SMDG1255415+191239), on the basis of a distance derived strictly from their recessional velocity, would fail the UDG physical size requirement. We refer to these galaxies as ``unconstrained.'' 
Four out of the seven UDG candidates with $1000 < cz_\text{CMB} < 2000$ km s$^{-1}$, also fall below the size requirement and are therefore non-UDGs. In summary, we have five galaxies (50\%) that we consider to be UDGs in this sample that have $cz_\text{CMB} < 4000$ km s$^{-1}$. 

Further characterizing our basic spectroscopic findings, we find that among candidates in our SMUDGes catalog for which we obtained LBT spectroscopic redshifts (19 targets), 15 are in the Coma region and four are in the off-Coma region. The Coma region can be roughly described as an area of projected radius of 
$\sim$19 Mpc ($11^\circ$) from the cluster center [$\alpha$=\ra{12}{59}{48.7}; $\delta$=\dec{27}{58}{50} (J2000)].
For our full sample of UDG candidates with spectroscopic redshifts, 68 in total, we exclude the 12 with $cz_\text{CMB} < 1000$ km s$^{-1}$, leaving us with 56 viable candidates. Among these, 44 (79\%) are UDGs by the relaxed surface brightness and size criteria we adopt, with all but one (SMDG0914401+283036) in the Coma-region, and 12 are non-UDGs (21\%).

\subsection{Environment Classification}
\label{environment_class}

To test whether environment plays a role in the formation and evolution of UDGs, we will be comparing the properties of UDGs that reside in different environments. To differentiate among possible environmental effects, we quantify the environment in two ways, one that is more sensitive to the global environment and another that is more sensitive to the local environment. For the global one, we define the environment based on membership in the Coma cluster, where that is designated using the UDG's relative position to the cluster in phase space, as described further below, and is meant to reflect whether 
the UDG has passed through the cluster and, having done so,
experienced the effects of the cluster environment. For the local one, we define the environment based on
the presence or absence of massive nearby galaxies, and is meant to reflect on whether galaxy-galaxy interactions play a role in determining UDG properties.

\medskip
\subsubsection{Defining Cluster Membership}
\label{cluster_mem}

To define cluster membership we opt for simple, rough criteria. We begin by taking a slice in recessional velocity that extends 2000 km s$^{-1}$ on either side of the Coma cluster mean recessional velocity of 7194 km s$^{-1}$ in CMB rest frame \citep[6925 km s$^{-1}$ in heliocentric rest frame;][]{Struble1999}. We then continue by making a cut in projected physical radius, $r_\mathrm{proj}$, requiring cluster members to be projected within Coma's splashback radius. We make this choice in radial cut guided by our science interest. We aim to limit our cluster sample to galaxies that could already have experienced the environmental effects of the cluster, as opposed, for example, to those within the virialized volume or those that are gravitationally bound to the cluster. We compute  $r_\mathrm{proj}$ for each object from the cluster center  at a redshift of $z=0.0231$. Using COLOSSUS \citep{colossus}, we find that Coma's splashback radius is 2.43 Mpc, adopting a virial radius of $r_{200}=1.99 \, h^{-1}$ Mpc and an enclosed mass of $M_{200}=1.88 \, h^{-1} \times 10^{15} \, \mathrm{M}_\odot$ \citep{Kubo2007}. These membership cuts are not as precise as those developed from the caustics visible in phase space \citep[see][]{Geller99}, but result in no material differences in our discussion.

Excluding the 10 nearby, unconstrained galaxies in the Coma-region and all 5 galaxies in the off-Coma region, we find that 40 of the 53 candidate UDGs (75\%) in the Coma region are within 2500 km s$^{-1}$ of Coma's mean recessional velocity and 13 (25\%) are either in the foreground or background. Of the 13 foreground and background objects, 9 (69\%) are UDGs with 5 in the foreground and 4 in the background.
Of those 40 UDG candidates within 2500 km s$^{-1}$ of Coma's mean velocity, 34 (85\%) are UDGs by our definition, 6 (15\%) are non-UDGs, and 9 (23\%) are UDGs that do not belong to the Coma cluster and which we consider to be field UDGs. 
Finally, we also include the one off-Coma UDG in the field UDG sample.

Due to the the direct adoption of Hubble flow distances, our subsequent analysis does not account for peculiar motion within the Coma cluster. We investigate the effects of peculiar motion on our analysis by assigning all cluster UDGs to the distance of the Coma cluster's mean recessional velocity in Section \ref{pecular_velocity}.

\medskip

\subsubsection{Defining Local Environment}

Next, we classify the local environment of the UDG candidates as sparse or dense.  We define as dense an environment with one or more massive companions ($M_g<-19$) within a projected separation of 300 kpc from the UDG candidate (calculated at the redshift of the candidate) and a $\Delta cz$ of $\pm$ 500 km s$^{-1}$. Otherwise, we categorize the environment as sparse. With these criteria, our sample of 44 UDGs (including SMDG0914401+283036 in the off-Coma region), divides into 21 UDGs in sparse environments and 23 in dense environments.

The environments designated for each UDG do not change when the velocity window is expanded to 1000 km s$^{-1}$, except in three cases (SMDG1301583+275011, 1256352+273507, and SMDG1300121+274823) for which the designation changes from sparse to dense. In our discussion we adopt the classifications using the smaller velocity window. As throughout, we exclude galaxies with $cz_\text{CMB}<1000$ km s$^{-1}$ from our discussion due to large distance uncertainties and designate the local environment of such objects as unconstrained.

\subsection{The Spatial Distribution of UDGs}

\begin{figure}[t]
\includegraphics[width=0.5\textwidth]{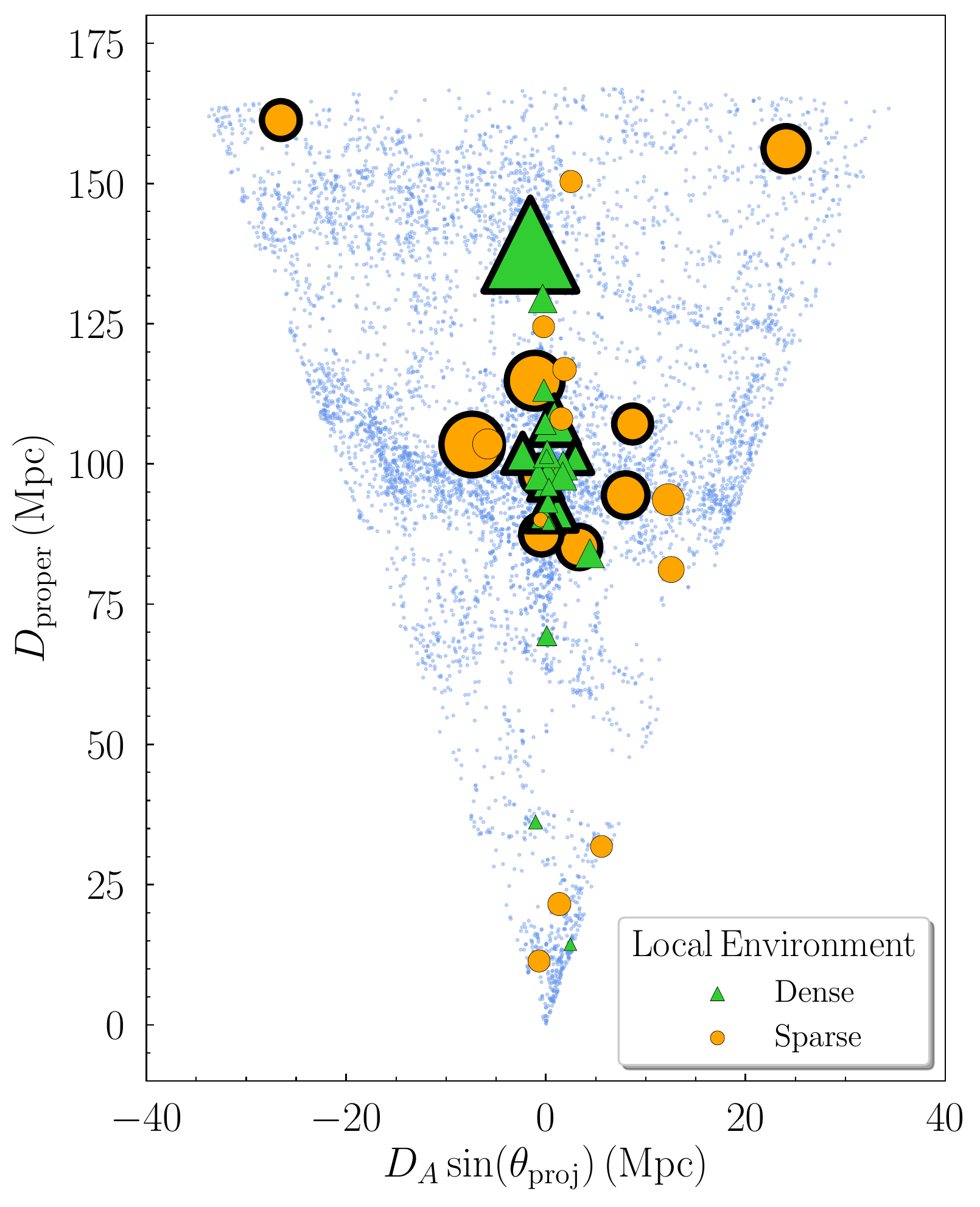}
\caption{The distribution of spectroscopically-confirmed UDGs relative to the large scale structure in the Coma region presented in a redshift wedge plot, depicting the physical projected separation, a product of the angular diameter distance $D_A$ at the redshift of the galaxy and the sine of the projected angular separation $\theta_\mathrm{proj}$, on the x-axis and the proper distance $D_\mathrm{proper}$ on the y-axis. Redshifts for high surface brightness galaxies are from SDSS, which are represented by blue dots. UDGs are coded by their local environment designations. The marker diameter linearly increases with the effective radius. Symbols have a bold outline if $r_e \geq 3.5$ kpc. For reference, the size of legend markers indicates the threshold size of UDGs ($r_e = 1.5$ kpc).}
\label{fig:zplot}
\end{figure}

In Figure \ref{fig:zplot}, we mark the positions of Coma-region, spectroscopically-confirmed UDGs in relation to the large scale structure in a redshift wedge plot with the physical projected separation, $D_A \sin(\theta_\text{proj})$, along the x-axis and the proper distance, $D_\mathrm{proper}$, along the y-axis. Most of the UDGs in our sample lie within the cluster or the filamentary structure associated with the Coma Cluster. Qualitatively, at least, the distribution of UDGs follows that of the high surface brightness on large scales. The larger UDGs appear to avoid the center of Coma, but we attribute this result, at least in part, to the selection of the sample. Confirmed UDGs within Coma come primarily from multi-object spectroscopic studies, which by construction mostly target typical UDGs, while UDGs outside of Coma come primarily from our own work that emphasized larger, rarer UDGs. This difference in the construction of the sample of UDGs inside and outside of Coma will, unfortunately, play a role in various aspects of our analysis and discussion.

As expected, there is tendency for UDGs in environments designated as locally dense to lie within the Coma cluster and as sparse to lie outside of it. However, there are exceptions, with a few UDGs in locally dense environments located outside of Coma. Such exceptions are potentially important test cases that can help us determine the relative importance of local vs. global environment on the properties of UDGs.  

\begin{figure}
\includegraphics[width=0.5\textwidth]{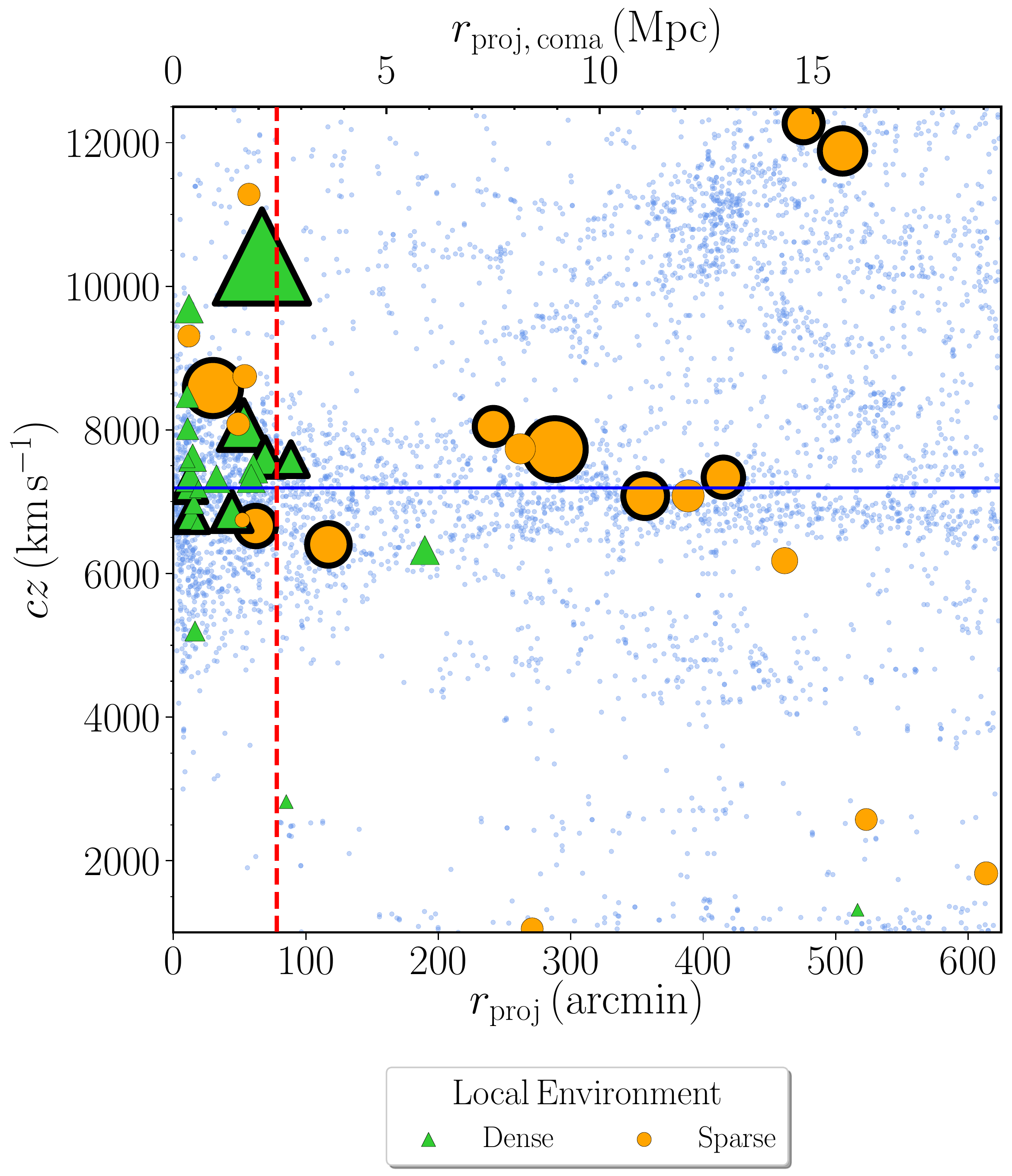}
\caption{Phase space diagram of the Coma cluster and near environs. The small blue dots represent individual, high surface brightness galaxies projected near the cluster, which mostly form the characteristic caustic pattern centered on Coma's mean recessional velocity (horizontal blue line). The dashed, red, vertical line marks Coma's splashback radius, estimated using the virial radius from \cite{Kubo2007} and COLOSSUS \citep{colossus}.
The green triangles and orange circles represent UDGs in dense and sparse environments, respectively. The markers indicating the locations of UDGs are outlined in black if the effective radius exceeds $r_e \geq 3.5$ kpc. The marker size linearly increases with the effective radius. For reference, the size of the legend markers indicates the threshold size of UDGs ($r_e = 1.5$ kpc). One UDG in our study fall off this plot;
SMDG0914401+283036 ($r_e=2.8$ kpc) does not lie in the region of the cluster. The physical projected radii (upper axis) are calculating assuming all objects lie on a tangentially projected plane at the distance of the Coma cluster.}
\label{fig:phase_space}
\end{figure}

In Figure \ref{fig:phase_space}, we present the phase-space distribution of UDGs relative to that of high surface brightness galaxies in and around the Coma cluster and compare their local environments. 18 of 24 Coma UDGs (75\%) reside in what we classify as locally dense environments. Among those UDGs, the ones that we classify as
residing in locally sparse environments tend to scatter farther in recessional 
velocity from the Coma mean velocity ($\sigma_{cz}$ = 1304 $\pm$ 210 km s$^{-1}$) than those that we
classify as residing in locally dense environments ($\sigma_{cz}$ = 892 $\pm$ 231 \, km s$^{-1}$), with only 
a 1.5\% random chance of such a difference. 
We speculate that the Coma UDG sample classified as residing in locally sparse environments is partly contaminated by foreground or background. 
Beyond the splashback radius, only 2 of 10 Coma-region UDGs (20\%) within $\pm 2500$ km s$^{-1}$ of the Coma cluster mean recessional velocity reside in a dense environment.
In conclusion, our two measures of environment highlight mostly the same galaxies as being in either high or low density environments. There are a few exceptions, but because of this nearly complete overlap we do not expect to reach vastly different conclusions depending on which measure of environment we choose to use. This resulting similarity in environment designations may just be a feature of the current sample, which is dominated by the Coma cluster. Future samples, where we have more systems in dense local environments that are not also in a dense global environment, may enable us to identify differences related to classes of dense environments. Such comparison have been illuminating in the study of high surface brightness galaxies \citep{Lewis2002,Gomez2003}.

\subsection{The Physical Size Distribution of UDGs}
\label{sec:size_distribution}

\begin{figure}
\includegraphics[width=0.5\textwidth]{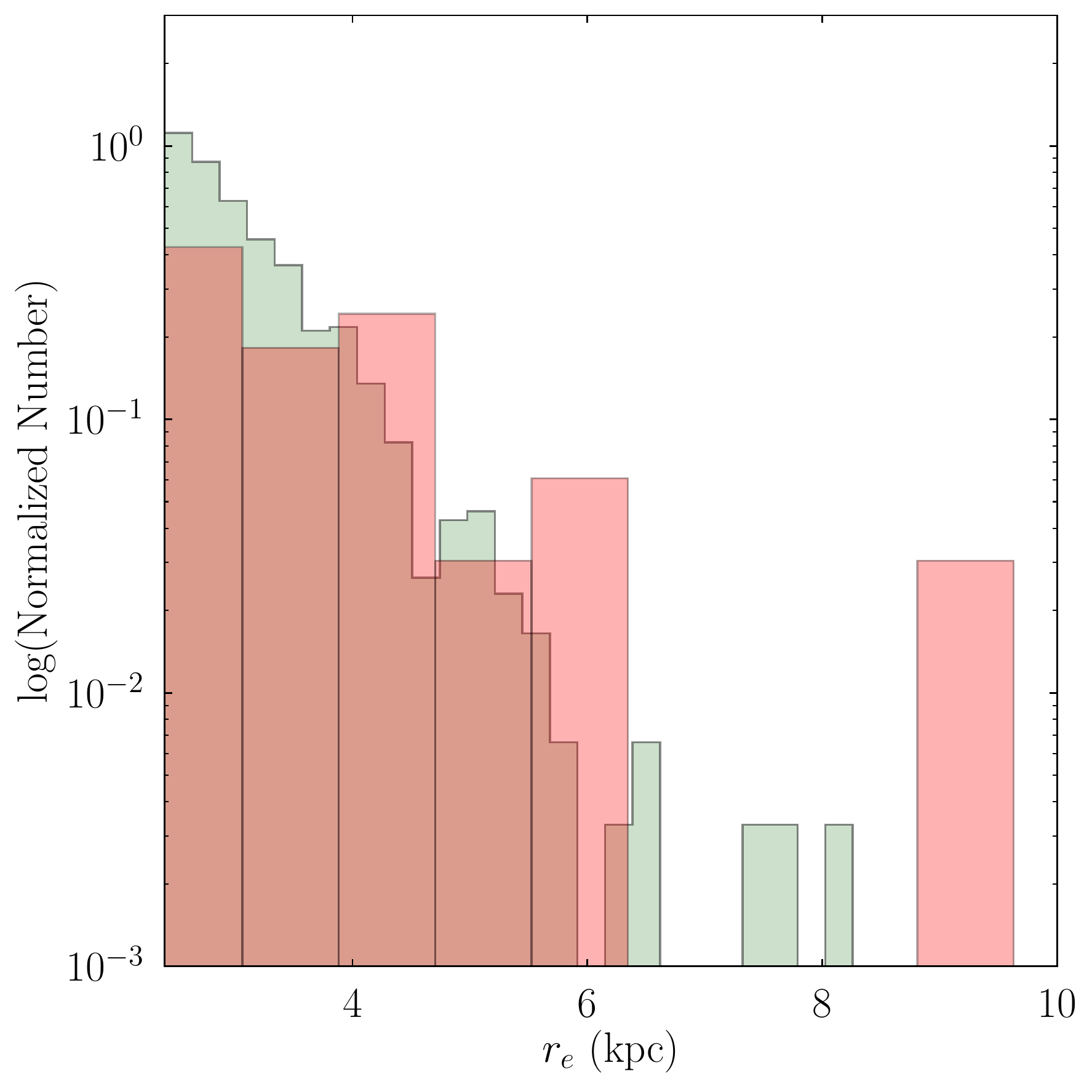}
\caption{Size histogram of UDGs (red) vs.\ SDSS galaxies (green) selected as described in text and normalized to equivalent numbers. UDGs and SDSS galaxies larger than 6 kpc are exceedingly rare.}
\label{fig:size_hist}
\end{figure}

Due to the observed correlation between the half-light radius and the derived enclosed mass (Figure \ref{fig:scaling}), finding the most massive UDGs requires searching among the largest UDGs. For this purpose, we consider UDGs with an effective radius $\geq 3.5$ kpc as potentially interesting. 
For example, DF44, (circularized $r_e = 3.9$ kpc), has an estimated halo mass of $\sim 10^{11}$ M$_\odot$ \citep{vanDokkum2019b,Saifollahi2020}.

These physically large systems are highlighted in Figures \ref{fig:zplot} and \ref{fig:phase_space} with bold, black outlined markers. We again urge caution in interpreting the distribution of these systems across the structure in Figure \ref{fig:phase_space} because different selection preferences exist among the spectroscopic studies. Namely, UDG redshift surveys using multi-object spectrographs target the core of the cluster to utilize the multiplexing capability. These surveys will include whatever candidates are in the field of view, large or small, while a long-slit spectroscopic survey such as ours, but also that of \cite{vanDokkum2015b}, will tend to focus on the larger, more physically interesting targets. Indeed, for this study we preferentially targeted systems with large angular extent across the various environments within the survey area. Nevertheless, 
physically large systems are found both inside and outside the cluster and in dense and sparse regions. As far as our environment designations can resolve, we do not find an environmental effect on the size of the largest UDGs.

As such, we now combine all of our UDGs and compare the size distribution of SDSS galaxies and UDGs in 
Figure \ref{fig:size_hist}. We have normalized the distributions for comparison and find that galaxies with $r_e \geq 6$ kpc are exceedingly rare (6 out of 1950 among the SDSS galaxies in this region and only two, SMDG1302166+285717 and SMDG1315427+311846, in our UDG sample). Although it is difficult to interpret any difference in the shapes of the two size distributions because of the biases described previously, the UDG size distribution is not strikingly different than that of the SDSS galaxies. 

Of the 44 UDGs in our sample, 16 have effective radii $\ge$ 3.5 kpc, including two exceptionally large ones that we just mentioned with $r_e > 6$ kpc (SMDG1315427+311846 with $r_e = 6.5$ kpc and SMDG1302166+285717 with $r_e = 9.9$ kpc).
The largest of these also has the smallest measured axis ratio ($b/a=0.37$) in our sample, which we suspect indicates that it is a tidal tail or tidally distorted galaxy rather than a UDG. 
Upon further examination, we find that it lies near a luminous early type galaxy (positioned slightly beyond the region shown in Figure \ref{fig:udgs}) that shares roughly the same redshift and that it is elongated in the direction toward this companion galaxy.
We will discuss the distribution of $b/a$ in our sample further in \S\ref{sec:axis_ratios}, but here we conclude that this source is unlikely to be a bona fide UDG. 

Using COLOSSUS \citep{colossus} and the \cite{Tinker2008} mass function, we estimate that halos of galaxies with $r_e >$ 6.0 kpc [which we estimate to correspond roughly to a halo mass of $10^{12.1} \, M_\odot$ (see Appendix) and up to a maximum $r_e$ of 10.0 kpc ($10^{12.5}) \, M_\odot$] comprise of 4\% of all galaxies larger than our cutoff of 1.5 kpc (with estimated corresponding mass of $10^{10.8} \, M_\odot$). Given that our sample has 44 UDGs, finding 1 or 2 UDG of such size is consistent with this expectation.
Given our preference for spectroscopically observing candidates of large angular extent, and the prevalence for the SMUDGes candidates in this area of sky to lie at roughly the Coma distance, we conclude from our results that UDG candidates with $r_e$ significantly larger than 6 kpc should be rare and viewed with some skepticism.

\begin{figure*}
\includegraphics[width=\textwidth]{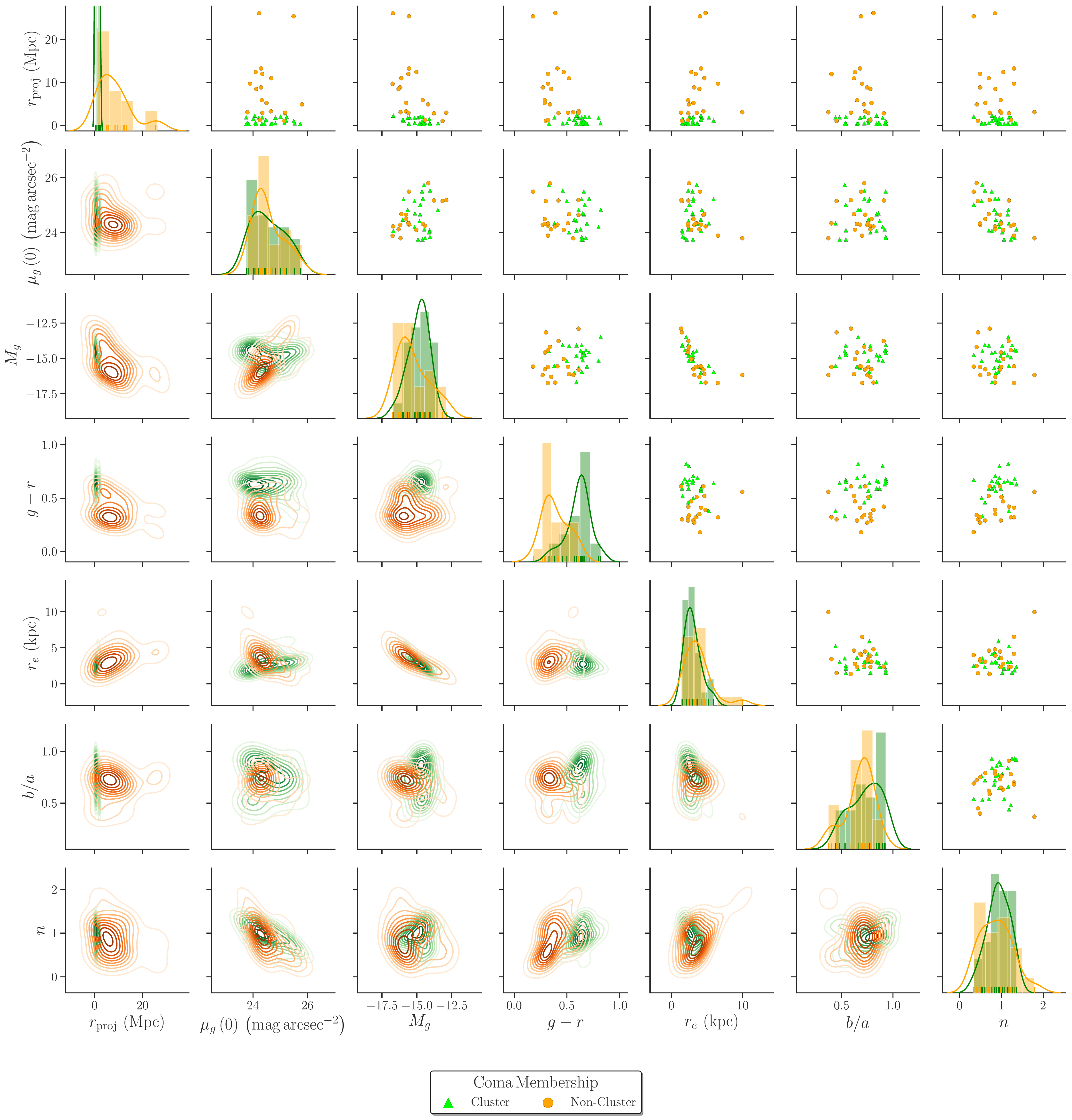}
\caption{Structural parameter distributions vs. cluster membership for UDGs. We present the distribution of the basic structural parameters for all spectroscopically confirmed UDGs in our sample. Coma cluster members and non-members are shown using different colors (green vs. orange, respectively). The upper triangular portion shows the individual data points, while the lower triangular portion shows the smoothed distribution. Panels along the diagonal show the marginalized distribution of the corresponding parameter. The projected, physical separation $r_\mathrm{proj}$ from the cluster center is computed at the distance of each UDG.}
\label{fig:pair_global_udgs}
\end{figure*}

\begin{figure*}
\includegraphics[width=\textwidth]{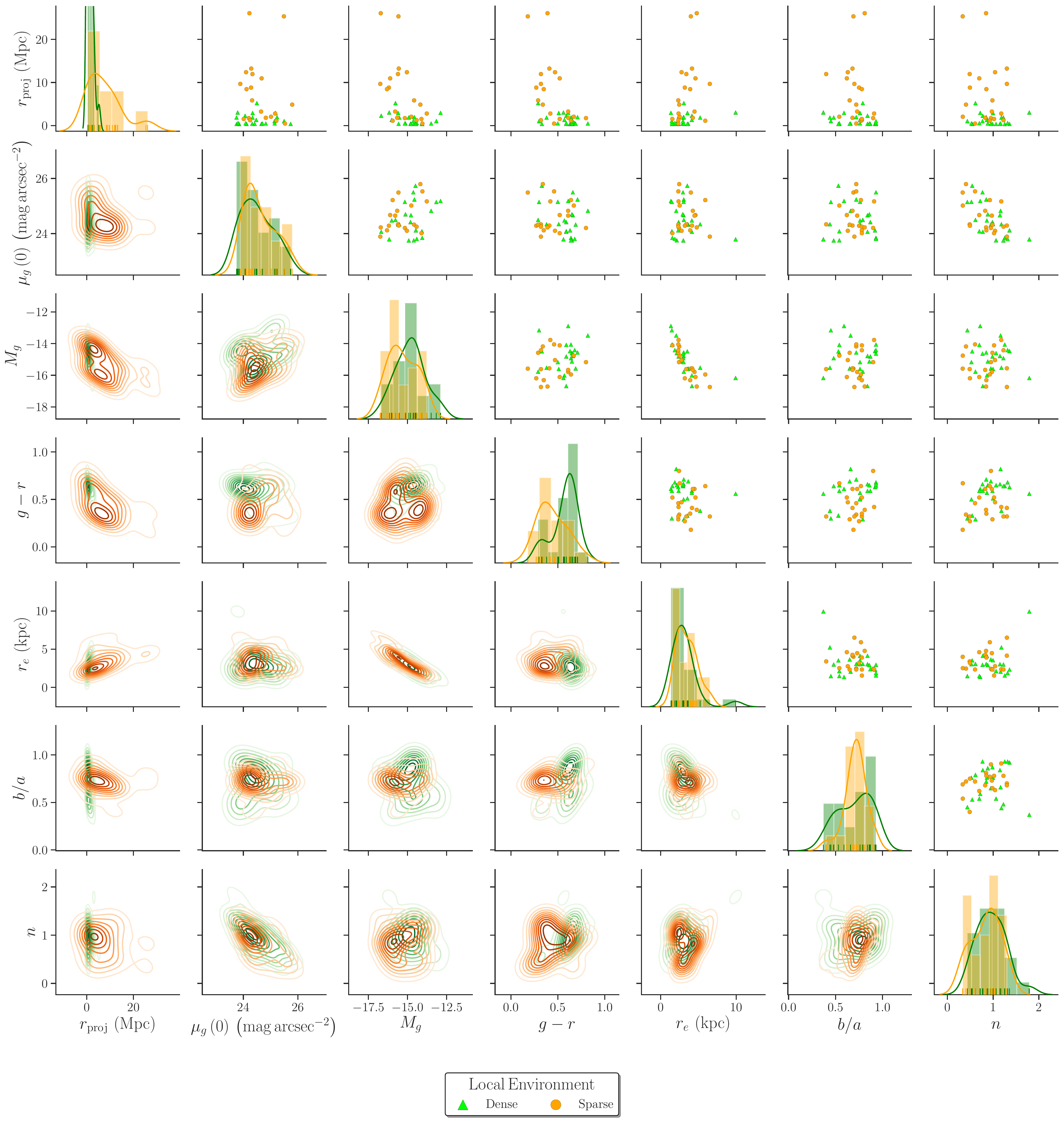}
\caption{Structural parameter distributions vs. local environment for UDGs. We present the distribution of the basic structural parameters for all spectroscopically confirmed UDGs in our sample. UDGs in dense and sparse local environments are shown using different colors (green vs. orange, respectively). The upper triangular portion shows the individual data points, while the lower triangular portion shows the smoothed distribution. Panels along the diagonal show the marginalized distribution of the corresponding parameter. The projected, physical separation $r_\mathrm{proj}$ from the cluster center is computed at the distance of each UDG.}
\label{fig:pair_local_udgs}
\end{figure*}

\subsection{Relationships Among UDG Structural and Photometric Parameters}

In this section we examine whether relationships exist among UDG structural and photometric parameters (i.e., surface brightness, luminosity, color index, effective radius, axis ratio, and Sersic index).
In Figures \ref{fig:pair_global_udgs} and \ref{fig:pair_local_udgs}, we present the data in graphical form, with UDGs distinguished according to global and local environments, respectively.
The contour plots represent the probability density of every unique pair of parameters and this is calculated using multivariate, kernel density estimation (KDE). The diagonal histograms feature the distribution of each parameter and its probability density.

We employ hypothesis testing to determine statistically significant feature correlations in UDGs. In addition, we identify features with underlying differences in their means and distributions in sparse and dense environments.
Specifically, for each pair of features analyzed in Figures \ref{fig:pair_global_udgs} and \ref{fig:pair_local_udgs}, we assess the statistical significance of a possible correlation using the Spearman's rank correlation test, the differences in the means using the independent sample Student's t-test, and in the distributions themselves using the Kolmogorov-Smirnov test. 
Because the probability of false positives increases when multiple inferences are made, we adopt the Bonferroni correction to compensate by requiring a higher confidence of rejection. To be specific, if we require a 5\% chance or less of reproducing the measured correlation in order to reject the hypothesis that the data are uncorrelated among a single set of parameters, we now set the requirement to $\alpha/m$, where $\alpha=5\%$ is the original significance level and $m=35$ is the total number of pairwise comparisons plus the number of comparisons made between feature mean and distribution shapes of UDGs.
We report only the statistically significant ($\ge$ 95\% confidence level) correlations and distribution differences under this heightened requirement

So defined, we find  statistically significant results in a limited number of tests. These break down into a few categories. First, the somewhat trivial cases include the correlation between size,  $r_e$, and luminosity, $M_g$. Also in this category are relationships between the projected radial separation from the Coma cluster and environmental density. 
Second are the cases that we suspect arise due to the different cluster and field selection. As we discussed previously, it is somewhat difficult to compare structural properties of UDGs inside and outside of the Coma cluster with this sample. This category contains the significant inverse correlation between separation from Coma and absolute magnitude. Finally, the third category consists of those trends that warrant further investigation and interpretation. In this category, we find evidence for correlations between
(1) central surface brightness, $\mu_g(0)$, and S\'ersic index, $n$; (2) $g-r$ and separation from the Coma cluster; and (3) for distinctions in the $g-r$ distributions with either cluster membership or environmental density.

\medskip
\subsubsection{$\mu_g(0)$ vs. $n$}

We observe a negative correlation between a UDG's central surface brightness and its S\'ersic index (Pearson correlation coefficient of $-0.59$, which has a chance of arising randomly of $7 \times 10^{-5}$ in this sample). Fainter galaxies have lower $n$. Higher signal-to-noise imaging is needed to determine whether there are two physical components playing off each other or whether there is a real range in the overall structure of a single component.
Overall, UDGs exhibit close to exponential surface profiles, with an average $n \sim 1$. The largest $n$ value in the sample presented here is 1.8 for  SMDG1302166+285717 (DF03). 

The trend we find between $\mu_g(0)$ and $n$ is consistent to that found among UDGs drawn from 8 galaxy clusters by \cite{Pina2019} and from Coma in both the Dragonfly \citep{vanDokkum2015a} and the Subaru \citep{Yagi2016} samples.
Furthermore, such a trend has also been
found among dwarf elliptical galaxies (dEs) in the Coma Cluster by \cite{Graham03}, 
who attributed it to the combination of what they interpret to be two more fundamental scaling relations: one between the magnitude $M_B$ and central surface brightness and the other between $M_B$ and  S\'ersic index. 
While we do not find significant evidence for these two trends in our UDG data,
when we look back at previous studies, we do find correlations between magnitude and $n$ and, in particular, between magnitude and surface brightness in both the \cite{Yagi2016} and \cite{Pina2019} samples, providing some support for the \cite{Graham03} postulate regarding the interplay of these three relations.

\medskip

\noindent
\subsubsection{$g-r$ vs. environment}
\begin{figure}
\includegraphics[width=0.5\textwidth]{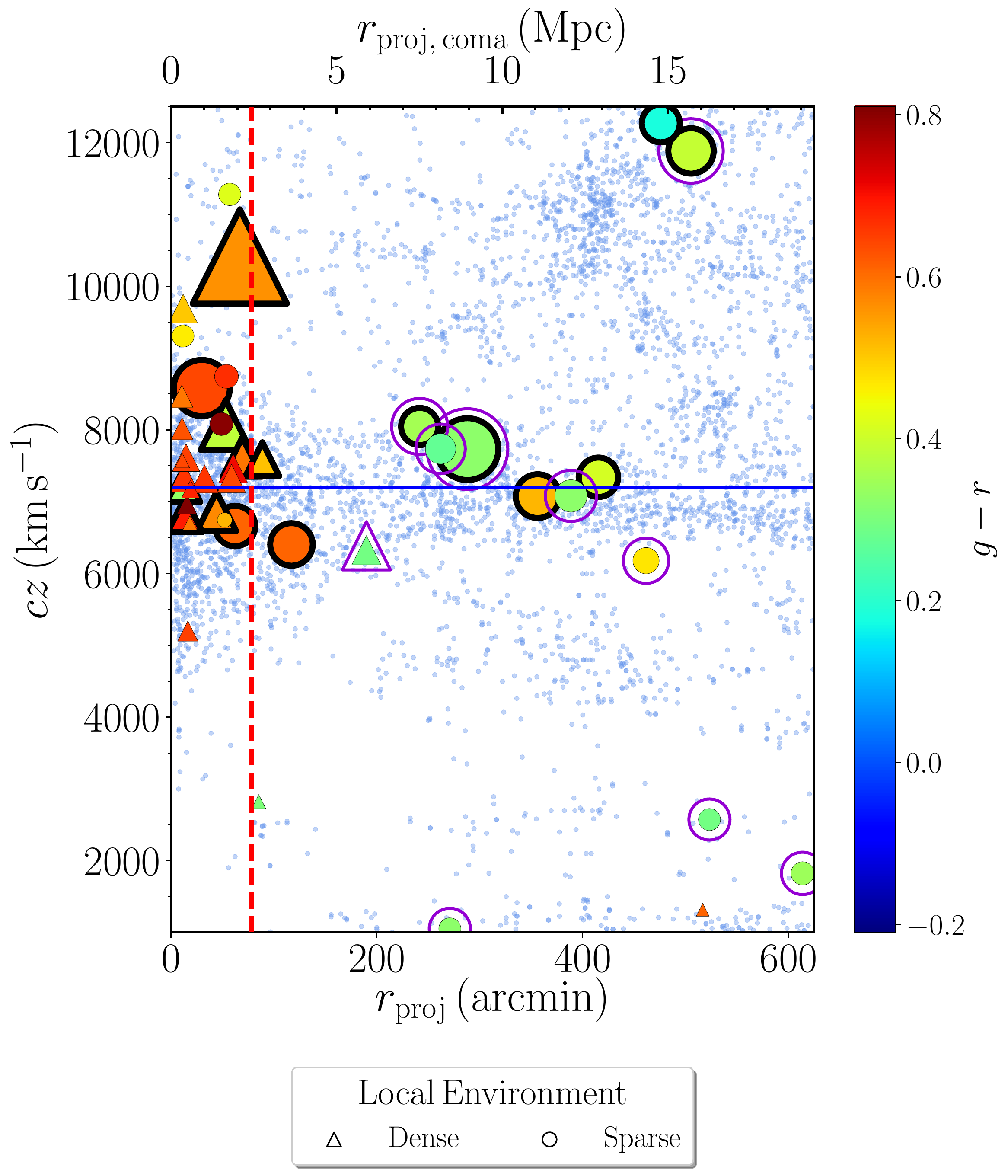}
\caption{Phase space diagram as in Figure \ref{fig:phase_space} with symbols coded by galaxy $g-r$ color. UDGs highlighted with a purple, concentric circles or triangles identify those with an NUV detection with signal-to-noise ratio greater than 3. An additional UDGs (SMDG0914401+283036) with significant NUV detections lies outside of the boundaries of the figure in the off-Coma region. The physical projected radii (upper axis) are calculating assuming all objects lie on a plane at the distance of the Coma cluster.}
\label{fig:udgs_local_phasespace_color}
\end{figure}

We observe a strong environmental-dependence on the UDG $g-r$ color. In Figure \ref{fig:udgs_local_phasespace_color}, we highlight the local environmental dependence of color within the Coma region phase-space diagram, with red UDGs predominantly residing in dense environments. From the perspective of the global environment, red UDGs are primarily cluster members. Given the strong correlation between our global and local environmental measures because we are so dominated by the Coma cluster, it is difficult to ascertain if there is a further dependence of star formation on local environment as has been shown to be the case for high surface brightness galaxies \citep{Lewis2002,Gomez2003}. However, as seen for the high surface brightness galaxies in those studies, the trend in diminished star formation for UDGs also appears to extend beyond the cluster's region of direct physical influence, whether that is interpreted to be the virial or splashback radius.

There is a potential concern that the color difference we identify between cluster and field (or high and low density environments) could be the result of heterogeneous spectroscopic selection and variable incompleteness in the absorption line spectroscopy of redder UDGs.
To address this concern, we examine results solely from our LBT sample. For this sample, the target selection is color-independent across environment and, although
it may be easier for us to measure a redshift for a bluer galaxy than for a redder one, we expect 
any possible color-dependent bias is also independent of 
environment. 
Despite the far smaller total sample size, we find 
a statistically significant rank correlation between UDG color and projected separation from the cluster center (a Spearman test rejects the null hypothesis with 96.7\% confidence).
The reality of the color trend with environment is further supported by the result that all 7 of our UDGs with GALEX detections, which are unbiased relative to Coma, are outside of the Coma cluster.

\begin{figure}
\includegraphics[width=0.5\textwidth]{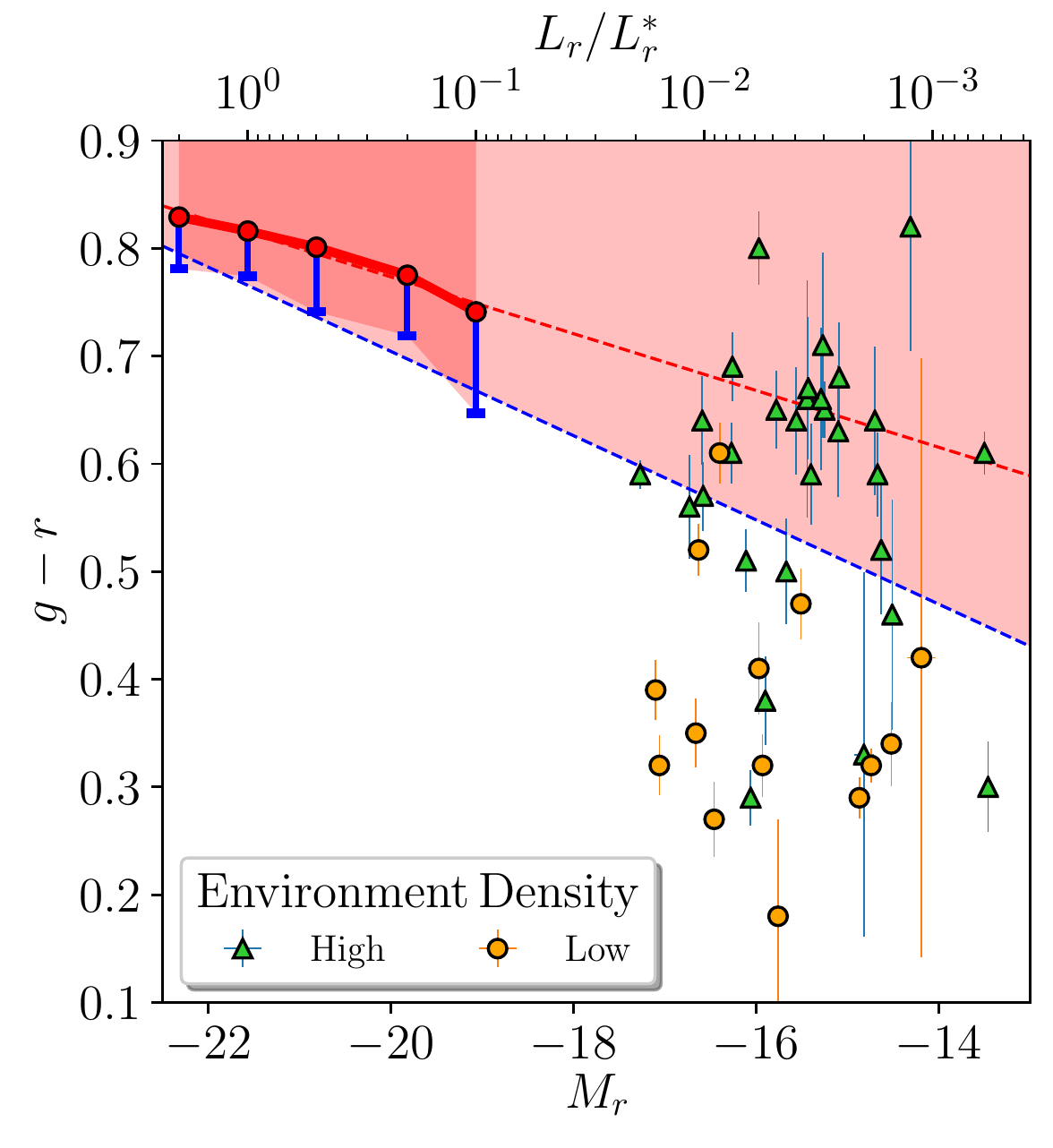}
\caption{Color-Magnitude Diagram of UDGs by environmental designation. We define the red sequence using the mean color of local elliptical galaxies from \cite{Schombert2016}, which are shown with open circles and downward error bars. These error bars represent twice the dispersion of individual elliptical galaxy colors. The dashed red and blue lines represent the extrapolation of the red sequence and 2$\sigma$ lower limit on the red sequence, respectively. The area above the 2$\sigma$ limit is shaded to highlight the region of the color-magnitude space consistent with early-type galaxies. We combine the local and global environment designations such that UDGs either in the cluster or in a locally dense environment are considered to be a high density environment. We find only one UDG in a low density environment that lies within the red sequence portion of the color-magnitude space.}
\label{fig:red_sequence}
\end{figure}

To probe further, we plot the color-magnitude diagram for UDGs in Figure \ref{fig:red_sequence} and compare to the location of the early-type red sequence galaxies.
We define the location of the early-type red sequence by linearly extrapolating the mean $g-r$ colors and their $2\sigma$ lower limit for elliptical galaxies \citep{Schombert2016} in 5  luminosity bins ($2.0 L_*$, $1.0 L_*$, $0.5L_*$, $0.2 L_*$, and $0.1 L_*$). We convert the luminosities to the r-band absolute magnitudes using the measured luminosity function of SDSS DR6 galaxies in the r-band \citep{Montero-Dorta}, which has a characteristic magnitude of $M_r^* - 5 \log_{10} h = -20.71 \pm 0.04$.
While the adoption of $z$-band absolute magnitude would better correlate with mass, we opt for the adoption of $r$-band due to the high sky brightness in $z$-band, which would yield less accurate photometry. Using $z$-band absolute magnitudes does not alter our conclusions from this analysis.
We consider all galaxies with $g-r$ colors greater than the extrapolated $2\sigma$ red sequence lower limit to be red. Given the similarity of the environmental behavior observed so far when environment is defined either globally or locally, we now merge the two designations. For this discussion and Figure \ref{fig:red_sequence}, we consider a UDG to be in a dense environment if it lies in either in the cluster environment or a locally dense one. From this extrapolated relation, we determine that all but one red UDG (SMDG1251014+274753) reside in high density environments. 

We extend this discussion by considering UV observations.
\cite{rs} present UV photometry of the \cite{Zaritsky19} catalog, providing limits in most cases but detections of 16 UDGs. Their analysis was conducted assuming that most of the sources are at the distance of the Coma cluster and now we evaluate whether that is likely to be the case. Unfortunately, among their detections we only match four with spectroscopic redshifts
(SMDG1237294+204442, SMDG1306148+275941, SMDG1313189+312452, and SMDG1315427+311846). We note that we do not replicate the NUV detection for SMDG1313189+312452 as the $S/N=2.7$ falls just below our detection threshold. Only one of these is consistent with lying at the distance of Coma and clearly satisfies the UDG criteria. Among those not detected in the UV, which is all of the other SMUDGes sources in Tables \ref{table:LBT_targets} and \ref{table:Non_LBT_targets1}, the fraction at the distance of Coma is  57\% (17/30). However, as we have noted before, our spectroscopic selection biases us toward foreground and so we cannot directly apply this result to the entire sample of candidates. Fortunately, most of the results presented by \cite{rs} are distance independent. 

To close this section, we
discuss three interesting red ($g-r \ge 0.5$) UDGs from 
Figure \ref{fig:udgs_local_phasespace_color}. These are (1) the large triangle projected within the splashback radius but relatively offset from the Coma mean recessional velocity (SMDG1302166+285717; DF03), (2) the ``red" UDG at $\sim$3.5 Mpc and 6000 km s$^{-1}$ (SMDG1251014+274753),  and (3) the ``orange" UDG at $\sim$ 10 Mpc and almost at the Coma recessional velocity (SMDG1240530+321655). The first of these is identified as lying in a dense local environment and we have already discussed this one as being a likely tidal tail or distorted object (\S\ref{sec:size_distribution}), so we disregard this galaxy. The second one is sufficiently close to the estimated splashback radius that, given uncertainties, perhaps it could have already made a pass through the Coma environment. So although this UDG is potentially an example of a passive UDG that is in a low density environment, it is not definitively so. The third UDG is this sample's best example of a quiescent field UDG. Its $g-r$ color places it just slightly below the red sequence region identified in Figure \ref{fig:red_sequence}, so perhaps it is not a truly passive galaxy. As such, we have found two interesting examples of potential passive field UDGs but no definitive ones and the majority of field UDGs are clearly either currently or recently star forming given their optical blue colors, and in many cases corresponding NUV detection.

\subsection{UDG Axis Ratios}
\label{sec:axis_ratios}

Given our findings that, aside from their colors, field and cluster UDGs are quite similar, if there is only a single UDG formation channel then it must be environment-independent.
Among the models proposed for the origin of UDGs, a subset are 
environment-independent. 
One such model stipulates that UDGs are galaxies that reside in high-angular momentum dark matter halos 
\citep{Amorisco2016}, while another suggests that baryonic matter in these galaxies retains a larger fraction of the their halo's angular momentum \citep{Posti2018}.
In either case, we expect a trend between the axis ratio, $b/a$, and effective radius, $r_e$, for UDGs of fixed stellar mass \citep[cf.,][]{Dalcanton1997}. 

\cite{Venhola2017}  investigated whether such a trend exists for UDGs and low surface brightness dwarf galaxies in the Coma and Fornax clusters. Their study, without considering the stellar masses of the UDGs, found a statistically significant anti-correlation, as expected in these models, between $b/a$ and $r_e$ for UDGs in the Fornax cluster but not for those in the Coma Cluster.

\cite{Pina2019} extended this type of analysis across 8 clusters, including 247 UDGs projected within $R_{200}$ of their respective host galaxy cluster and 195 UDGs lying in regions beyond this projected distance.
When considering all of their galaxies, binned by stellar mass, they find weak evidence for an anti-correlation between $b/a$ and $r_e$ for UDGs lying within $R_{200}$. They find no such trend for UDGs beyond $R_{200}$. 
\cite{Pina2019} conjectured that the UDGs may become more elongated after interacting within their host cluster, complicating any interpretation relating $b/a$ to the UDG's angular momentum.

We revisit this topic with our sample. Because we established that $r_e$ correlates strongly with the luminosity of the UDG, we must correct for this trend prior to comparing $r_e$ values among our UDGs.
Our concern is that because there is a mass (and luminosity) difference as a function of cluster-centric radius due to the selection of the samples,  a real relation between b/a and size might artificially suggest a false relation between b/a and cluster-centric radius (i.e., an environmental reason for changes in b/a). As such, we focus our investigation on whether UDGs that deviate from the mean $r_e$ for galaxies of their mass have unusual $b/a$. In particular, do UDGs with unusually large values of $r_e$, which might be due to tidal effects, also have unusually low values of $b/a$?
To standardize $r_e$, we adopt a functional form such that $M_r$ and $\log(r_e)$ are linear, because $M_r$ is linear with $\mu_r(0) + \log(r_e^2)$ for a S\'ersic profile.
We first measure the empirical relation between $r_e$ and $M_r$ using least squares regression and find,

\begin{equation}
r_e = \left(0.012 \cdot 10^{-0.15 M_r} \right) \, \mathrm{kpc}.
\end{equation}
\noindent
We then define the standardized effective radius, $r^\prime_e$, as the ratio between the measured $r_e$ and the expected value for the corresponding $\mathrm{M}_r$.  

\begin{figure}[t]
\includegraphics[width=0.5\textwidth]{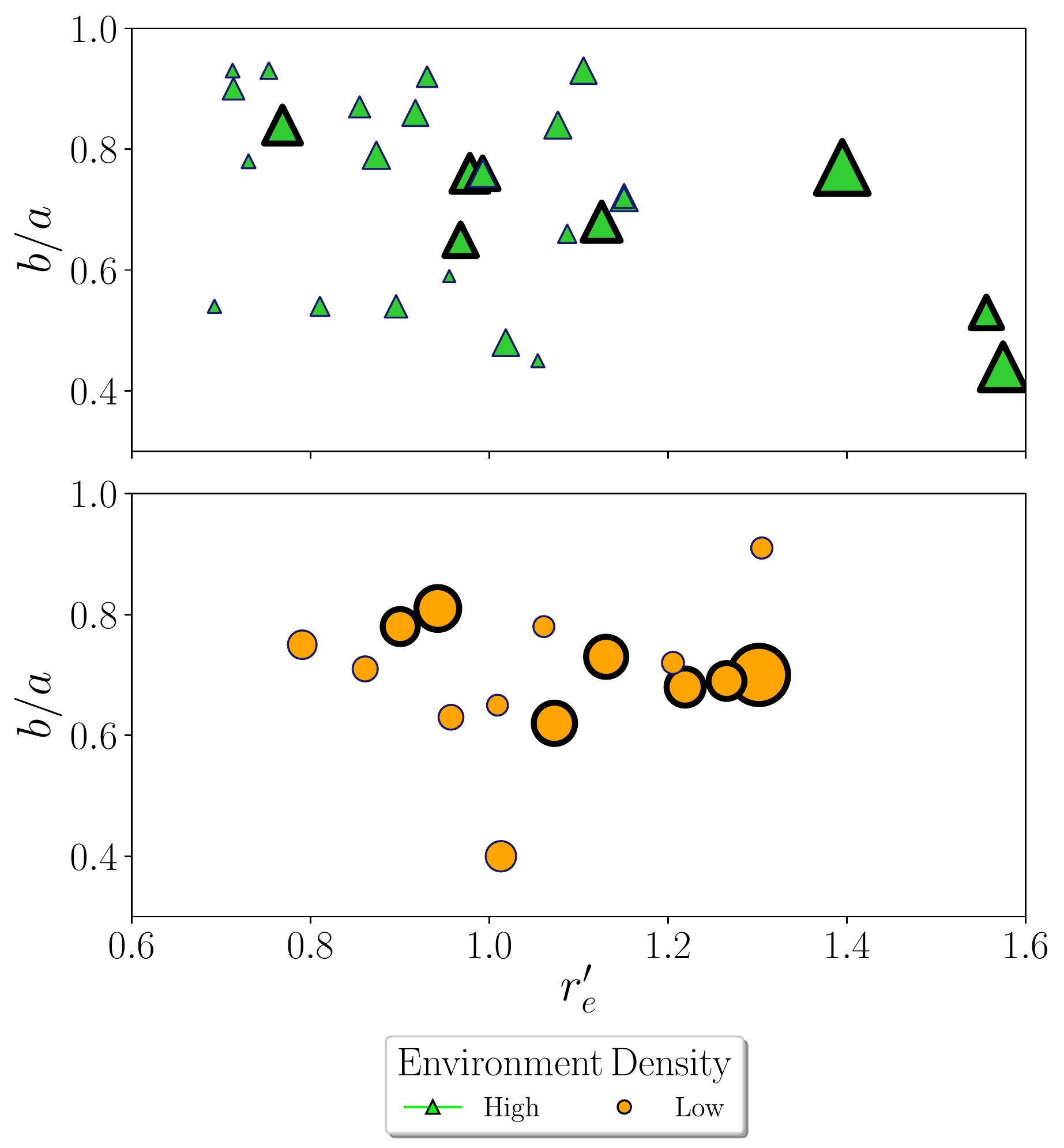}
\caption{Projected axis ratio,  $b/a$, vs.\ the standardized effective radius $r'_e$ of UDGs, separated by environment. The standardized effective radius removes the correlation between $r_e$ and absolute magnitude, $\mathrm{M}_r$. We find a hint of an anti-correlation between $b/a$ and $r^\prime_e$ (with a $0.032$ probability for occurring by chance) 
among UDGs in high density environments. We find no such trend between these two parameters among UDGs in low density environments.
Once again, the marker diameter increases linearly with the effective radius of the UDG and we highlight large UDGs (spanning $r_e \geq 3.5$ kpc) with bold outlines. For reference, the size of the legend markers correspond to UDGs with $r_e=1.5$ kpc.} 
\label{fig:ar_vs_re}
\end{figure}

We analyze UDGs in high and low density environments separately because of the environmental differences observed by \cite{Pina2019}.
In Figure \ref{fig:ar_vs_re}, we present the relation between $b/a$ and $r^\prime_e$.
Although there is a large scatter among UDGs in both environments, 
we find a hint of an anti-correlation between $b/a$ and $r^\prime_e$ for UDGs in high density environments (chance of random occurrence is only 0.032).
We find that there is no significant correlation for UDGs in low density environments (chance of random occurrence is 0.25).
These results are in agreement with those found by \cite{Pina2019}.

We caution that it is premature to use these results to bolster or refute certain models that make predictions for the trends in $b/a$ \citep[i.e,][]{Amorisco2016, Posti2018}. The scatter in Figure \ref{fig:ar_vs_re} is large and our data limited.
There may also be results imposed by the sample selection. For example, consider that selecting galaxies with a highly limited range of central surface brightness imprints a relationship between galaxy flux and projected size. If we select for constant flux (or luminosity or stellar mass assuming all galaxies are at the same distance), axis ratios will need to decrease with increasing major axis, approximated by the effective radius, to maintain a constant projected area.
Alternatively, rounder objects of the same effective radius will have a larger total flux. 
Although the UDG central surface brightness range in photometric samples is $\sim2$ magnitudes, the range in the spectroscopic studies is much more limited due to practical observing constraints.
In another example of potential complications, the spectroscopic sample is likely to be affected by non-uniform $b/a$ selection criteria among studies, such as the rejection of UDGs with smaller axis ratios in one or another study on the basis that these candidates might more likely be interlopers.
Finally, because blue UDG candidates are typically clumpy and red ones are typically smooth \citep{karunakaran20}, fitting smooth GALFIT models may result in different systematics. For example, \cite{karunakaran20} argue that the application of smooth GALFIT models to clumpy star-forming UDGs systematically underestimates the axial ratio because the fit parameters are being pulled by the clumps instead of the underlying, fainter light from the smoother, older population. While their hypothesis needs to be tested with simulations, the sense of differences between cluster and non-cluster UDG axial ratios in Figure \ref{fig:ar_vs_re} is consistent with that systematic effect. All of these issues must be satisfactorily resolved before venturing to use $b/a$ to evaluate theoretical models.

\begin{figure}[t]
\includegraphics[width=0.5\textwidth]{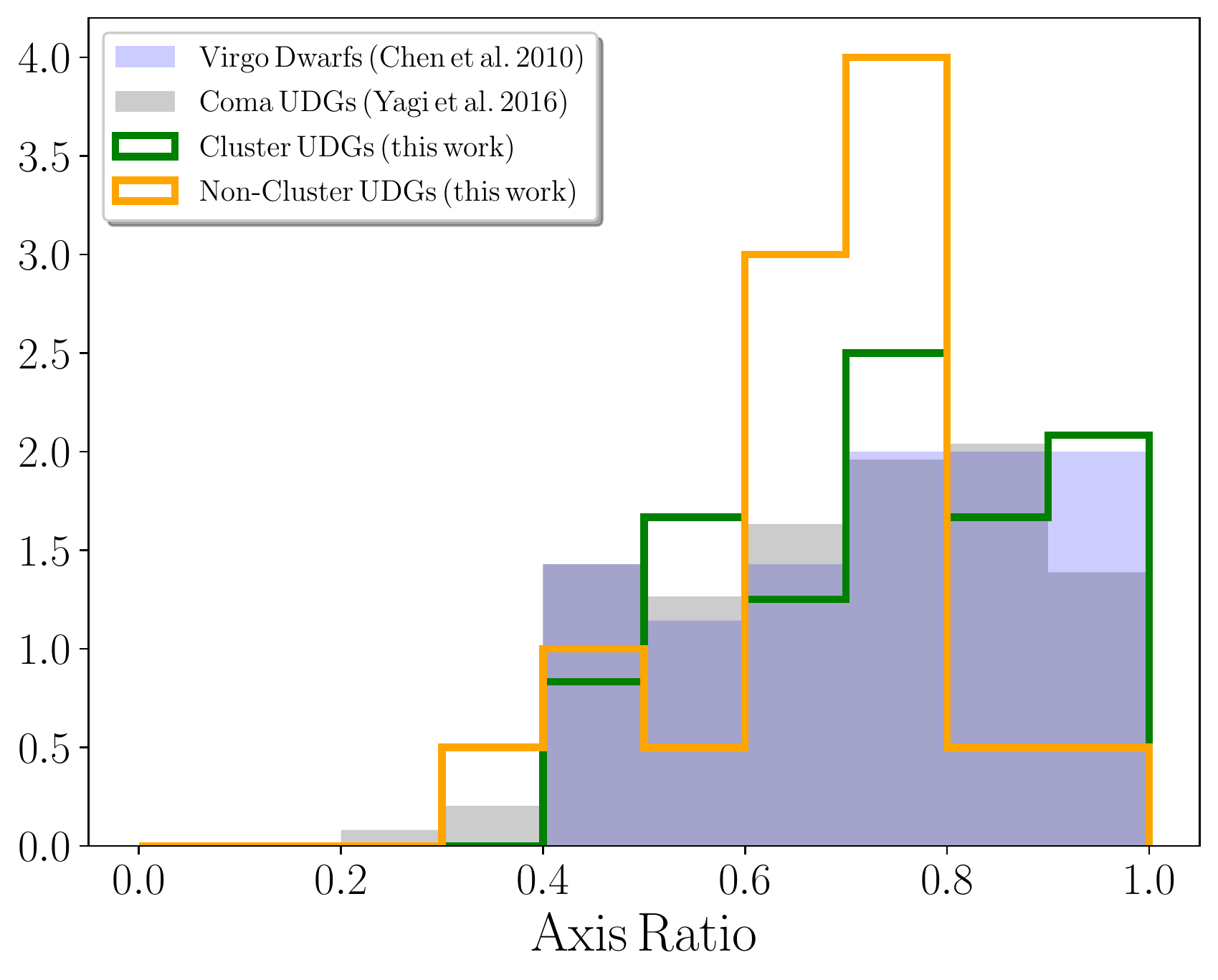}
\caption{Axis ratio distributions of cluster (green outline) and non-cluster (orange outline) UDGs from this work, Virgo dwarf galaxies (light blue shade) from \cite{Chen2010}, and cluster UDGs (gray shade) from \cite{Yagi2016}. The same effective radius ($r_e \geq 1.4$ kpc) and surface brightness ($\mu_g(0) \geq 23.8$ mag arcsec$^{-2}$) cut offs have been applied to the \cite{Yagi2016} sample to match our distribution. The histogram of each population is normalized to have an area of 1.}
\label{fig:axis_ratio}
\end{figure}

Independent of UDG formation models, 
the results shown in Figure \ref{fig:ar_vs_re} suggest a distinction in the axis ratio properties of UDGs in dense and sparse environments. As such, 
the geometry of UDGs might provide insightful clues to their likely formation and evolutionary pathways beyond the expectation drawn from models involving angular momenta. To examine further the effects of the environment on the evolution of these enigmatic galaxies, we present the UDG axis ratio distribution inside and outside the Coma cluster. We find a relatively flat distribution of projected axial ratios from 0.4 to 1 for cluster UDGs, as shown in Figure \ref{fig:axis_ratio}.
The absence of galaxies with $b/a < 0.4$ is at least partially due to SMUDGes selection criteria \citep{Zaritsky19}.
This flat distribution is consistent with findings from larger studies of cluster UDGs \citep{Koda2015, Burkert2017} and dwarf elliptical galaxies \citep{Chen2010, Lisker2009}.

We measure a mean projected axis ratio that is consistent between our cluster and non-cluster UDG populations (0.73 and 0.69, respectively) and with previous studies of similar populations. The mean axis ratio of Coma Cluster UDGs from the \cite{Yagi2016} catalog varies between 0.69 to 0.72, depending on the selection criteria used for each study \citep{Koda2015, Yagi2016, Burkert2017}. The low-surface brightness galaxy population in \cite{Chen2010} has a mean axis ratio of 0.73 $\pm$ 0.18. We perform the Student's t-test to compare the mean of our non-cluster population with that of the cluster population from the \cite{Yagi2016} catalog with effective radius and surface brightness cuts to match those used in our study ($r_e\geq 1.4$ kpc and $\mu_g(0) \geq 23.8$). We find no significant difference between these samples. Despite the lack of statistical evidence for differences, there are hints of potential differences in the distribution of values for field and cluster UDGs. First, we do not find the same absence of round UDGs ($b/a = 0.9 - 1.0$) as found by \cite{Burkert2017}, which had supported their conclusion that Coma Cluster UDGs favor a prolate geometry.  Second,
the distribution of apparent axial ratios for our cluster and non-cluster populations are visually quite different (Figure \ref{fig:axis_ratio}). However, when we conduct the Kolmogorov-Smirnov and Anderson-Darling tests to assess this possibility, the differences do not rise to a sufficiently high confidence level, presumably because of the small sample sizes involved. We conclude that there are potentially interesting signatures that need to be explored with larger samples.

\subsection{Effects of Local Peculiar Velocity}
\label{pecular_velocity}

To assess the effects of peculiar velocity due to the Coma cluster itself to our distance determinations, which we ignore in our initial treatment,  we adopt the alternative extreme possibility and place 
all cluster UDGs at the distance corresponding to Coma's mean recessional velocity (7194 km s$^{-1}$ in CMB-rest frame).
This change results in small differences in the measurements of the UDGs’ effective radii, magnitudes, colors, and projected separations from the Coma Cluster, while the UDGs' surface brightness ($\mu_g(0)$), axis ratios ($b/a$), and S\'{e}rsic indices ($n$) remain unchanged. 

While our initial analysis yielded a relatively even split between UDGs in sparse (21; 48\%) and dense environments (23; 52\%), our new distance designations guarantee that all cluster UDGs are in a locally dense environment. The resulting change in the local environment assignment creates a skewed split with 15 UDGs (34\%) in sparse and 29 UDGs (66\%) in dense environments. The most significant results arising from the updated physical parameters and local environment assignments is a more pronounced $g-r$ color difference between locally dense and sparse environments.
 
When redoing the correlation analysis, we are able to statistically reject the same hypotheses as we did previously.
In addition, we are able to statistically reject one additional null hypothesis (with a 0.05\% random chance of arising) and find support for a correlation between the size of the UDG ($r_e$) and its projected separation from Coma ($r_\mathrm{proj}$). As discussed previously however, such correlations are likely due to the different spectroscopic target selection between the literature samples and our own. 
The cluster studies favor the typical (i.e., smaller) UDGs located within the field of view of a multi-object spectrograph, where the field study (ours) targeted the largest UDG candidates across environments.


\section{Conclusions}

From our spectroscopic follow-up campaign of SMUDGes candidate UDGs, we reach a set of milestones and conclusions. 

\begin{enumerate}
    \item We confirm two key observational challenges in the study of UDGs. First, even a significant exposure time ($\sim1$ hour) on 10-m class telescopes often fails to yield a redshift for a candidate UDG (in our case 
    for nearly half of our targets). Because this failure can, at least in part, depend on the lack of strong absorption lines, there is the possibility that final spectroscopic samples are biased in terms of age, metallicity, or star formation history. Second, field samples of UDG candidates can be significantly contaminated by non-UDGs. In our case, the percentage of non-UDGs, once we establish physical sizes, was $\sim$ 40\%. 
    
    \item Despite these difficulties, after combining our results with those in the literature for UDGs in the Coma cluster, we compiled a sample of 44 spectroscopically-confirmed UDGs with which to explore the nature of UDGs in and around the Coma cluster.
    
    \item Outside of the Coma cluster, the confirmed UDGs divide roughly evenly between those lying within the large scale structure surrounding Coma and those completely unassociated with Coma. We classify all of these as field galaxies for our discussion. All but five of these (15 of 20) are in what we categorize as sparse environments (i.e. not projected within 300 kpc of a massive galaxy ($M_g < 19.0$) with a relative velocity difference $<500$ km s$^{-1}$).
   
    \item Large UDGs ($r_e > 3.5$ kpc) are found both in the cluster and field environment. We find no measurable dependence of size on environment, but the cluster and field samples are selected differently which complicates any interpretation of the current sample.
    
    \item We conclude that the largest UDG in our sample (SMDG1302166+285717, $r_e = 9.9$ kpc) is likely to be the result of tides (either a tidal filament or tidally distorted) due to its unusually large size, its extreme axis ratio, its association with a nearby massive galaxy, and its orientation towards that same galaxy.
    
    \item Excluding that one galaxy, we find that the distribution of UDG sizes is similar to that of SDSS galaxies, with $r_e > 6$ kpc  galaxies being exceedingly rare. This result coincides well with our estimated halo mass for such a UDG of $\sim 10^{12}$ M$_\odot$. Galaxies of this mass or greater should make up roughly 4\% of all our UDGs, consistent with our finding 1 out of 43. 
    
    \item We confirm the earlier findings of a color-local density relation for UDGs \citep{greco,tanog} with the statistically strong result that UDGs in low density environments are highly unlikely to lie on the galaxy red sequence (only 1 out of 15 in our sample). Furthermore, we find that UDGs in high density environments are mostly red (20 out of 29 in our sample), which in reality is likely to be an even stronger effect because some systems may be projected onto the high density environment of the Coma cluster. 
    
    \item We find two potentially passive, large UDGs in the field, suggesting that the cluster environment is not required to quench star formation. The sample is small; one of the two UDGs may be sufficiently close to Coma to have experienced the cluster environment and the optical color of the second is only marginally consistent with it being on the galaxy red sequence. More such objects must be identified to confirm this finding and establish whether passive, field UDGs exist.
   
    \item We detect NUV flux from a large fraction of the UDGs in sparse environments (10 out of the 20 in our sample for which we have {\sl GALEX} data). This result is consistent with the results of \cite{greco} and we interpret it to mean that recent or ongoing star formation is common among field UDGs and likely to be episodic.
   
   \item 
    While previous findings \citep{Pina2019} showed a weak relation between axis ratio ($b/a$) and size ($r_e$) for UDGs in the cluster environment and a lack of any trend between the two quantities for UDGs outside of the cluster environment, we confirm similar results using standardized $r'_e$ in lieu of $r_e$. For various reasons that we discuss, however, we recommend against physical interpretations of this result at the current time.
    
    \item We find no statistically significant difference in the axis ratios of UDGs in the Coma cluster and field, although there is a suggestion that the field UDGs have a narrower distribution of $b/a$. This result merits subsequent study with larger samples.

\end{enumerate}

Spectroscopic observations of UDG candidates provides necessary confirmation of the UDG's physical characteristics, and they are particularly important for establishing candidates in low density environments where association cannot be used to assign a distance. We will continue observationally expensive programs in the optical, such as reported here, and in the radio \citep{karunakaran20} to supplement indirect distance constraints on UDGs. 

\acknowledgments
JK, DZ, and RD acknowledge financial support from NSF AST-1713841 and  AST-2006785. KS acknowledges support from the Natural Sciences and Engineering Research Council of Canada (NSERC).
We made use of the online cosmology calculator as described in \cite{cosmo_calc}
and thank Barry Rothberg and Olga Kuhn, our LBT support astronomers, for their help during our observation runs. In particular, Olga Kuhn helped us to navigate through the MODS data reduction pipeline. In addition, we like to thank Wen-fai Fong who took our calibration data for our Feb 2017 observation run. And lastly, we thank the anonymous referee for the thorough comments and suggestions, which has improve our manuscript.

The LBT is an international collaboration among institutions in the United States, Italy and Germany. LBT Corporation partners are: The University of Arizona on behalf of the Arizona university system; Instituto Nazionale di Astrofisica, Italy; LBT Beteiligungsgesellschaft, Germany, representing the Max-Planck Society, the Astrophysical Institute Potsdam, and Heidelberg University; The Ohio State University, and The Research Corporation, on behalf of The University of Notre Dame, University of Minnesota and University of Virginia.
This paper uses data taken with the MODS spectrographs built with funding from NSF grant AST-9987045 and the NSF Telescope System Instrumentation Program (TSIP), with additional funds from the Ohio Board of Regents and the Ohio State University Office of Research.

The authors made use of the NASA/IPAC Extragalactic Database (NED) which is operated by the Jet Propulsion Laboratory, California Institute of Technology, under contract with the National Aeronautics and Space Administration.

\facility{LBT (MODS); GALEX}

\software{
Astropy 3.2.1        \citep{astropy1, astropy2},
COLOSSUS 1.2.11      \citep{colossus},
IPython 4.0.0        \citep{ipython},
Matplotlib 3.1.0     \citep{matplotlib},
NumPy 1.16.4         \citep{numpy1, numpy2},
pandas 0.25.1        \citep{pandas},
scikit-learn 0.21.2  \citep{sklearn},
SciPy 1.3.0          \citep{scipy1, scipy2}
seaborn 0.9.0        \citep{seaborn},
Cosmology Calculator \citep{cosmo_calc},
GALFIT 3.0.5         \citep{galfit1, galfit2},
IRAF 2.16            \citep{iraf1, iraf2},
L.A. Cosmic          \citep{vanDokkum2001},
MODS Data Reduction Pipeline 1.3p1 \citep{Pogge2010}
}

\newpage
\begin{appendix}
\renewcommand\thefigure{\thesection.\arabic{figure}}  
\setcounter{figure}{0}

\section{Size vs. Mass}
\label{appendix}

We have focused our spectroscopic observations on UDGs from the SMUDGes catalog that have large angular extents. We do this in an attempt to select those that are the physically largest as well. As our redshifts show, we have a mixed selection of objects in the end, including some nearby, small galaxies. Here we motivate, in somewhat more detail, why we aim to study the physically largest UDGs. 

\cite{Zaritsky2017} suggested, on the basis of the two UDGs with velocity dispersion measurements at the time, that UDGs satisfy a scaling relation satisfied by other galaxies \citep[referred, in reference to its antecedent, to as Fundamental Manifold or FM]{Zaritsky06,Zaritsky08}. By using the FM relation to estimate the velocity dispersions of other UDGs, that study suggested that UDGs span a range of masses and that effective radius correlates with halo mass. The latter is a result of FM relation and the fact that UDGs have roughly the same surface brightness due to how they are selected.

Here we revisit the earlier finding by examining the much larger set of UDGs that currently have a measured velocity dispersion. As before, the sample is still a mix between systems with integrated stellar dispersions and dispersions measured using individual globular cluster velocities. 
We adopt velocity dispersions from a range of sources \citep{Beasley2016b,vanDokkum2017,toloba,Chilingarian2019,mn,vanDokkum2019b} and structural parameters from references therein. With a few exceptions that we now describe, we have used the data as presented and include all known UDGs with velocity dispersion measurements. For DF44, there are conflicting results \citep{vanDokkum2017,vanDokkum2019b} that are not understood. The original \cite{vanDokkum2017} dispersion measurement was modified by \cite{vanDokkum2019b} after a modest error was identified in the original treatment, but the revised value is still discrepant within the internal uncertainties with their new value. We opt to average these two values and adopt the larger of the presented uncertainties.
From the set of Virgo UDGs presented by \cite{toloba}, we exclude VLSB-D which has $r_e = 13.4\pm0.2$ kpc. This scale length is much larger than any other UDG. We conclude that it is a tidal feature rather than a galaxy, as we did with our largest candidate as well.
Lastly, we have excluded two UDGs near NGC 1052 with measured velocity dispersions \citep[NGC1052-DF2, NGC1052-DF4;][]{Danieli2019,vanDokkum2019a} because of their controversial nature and suggestions that they may be dark matter free and therefore quite different than the other systems \citep{vanDokkum2018,trujillo2019,danieli20}. We discuss farther below why their exclusion does not qualitatively affect our conclusion.

\begin{figure*}[t]
\center
\includegraphics[width=0.5\textwidth]{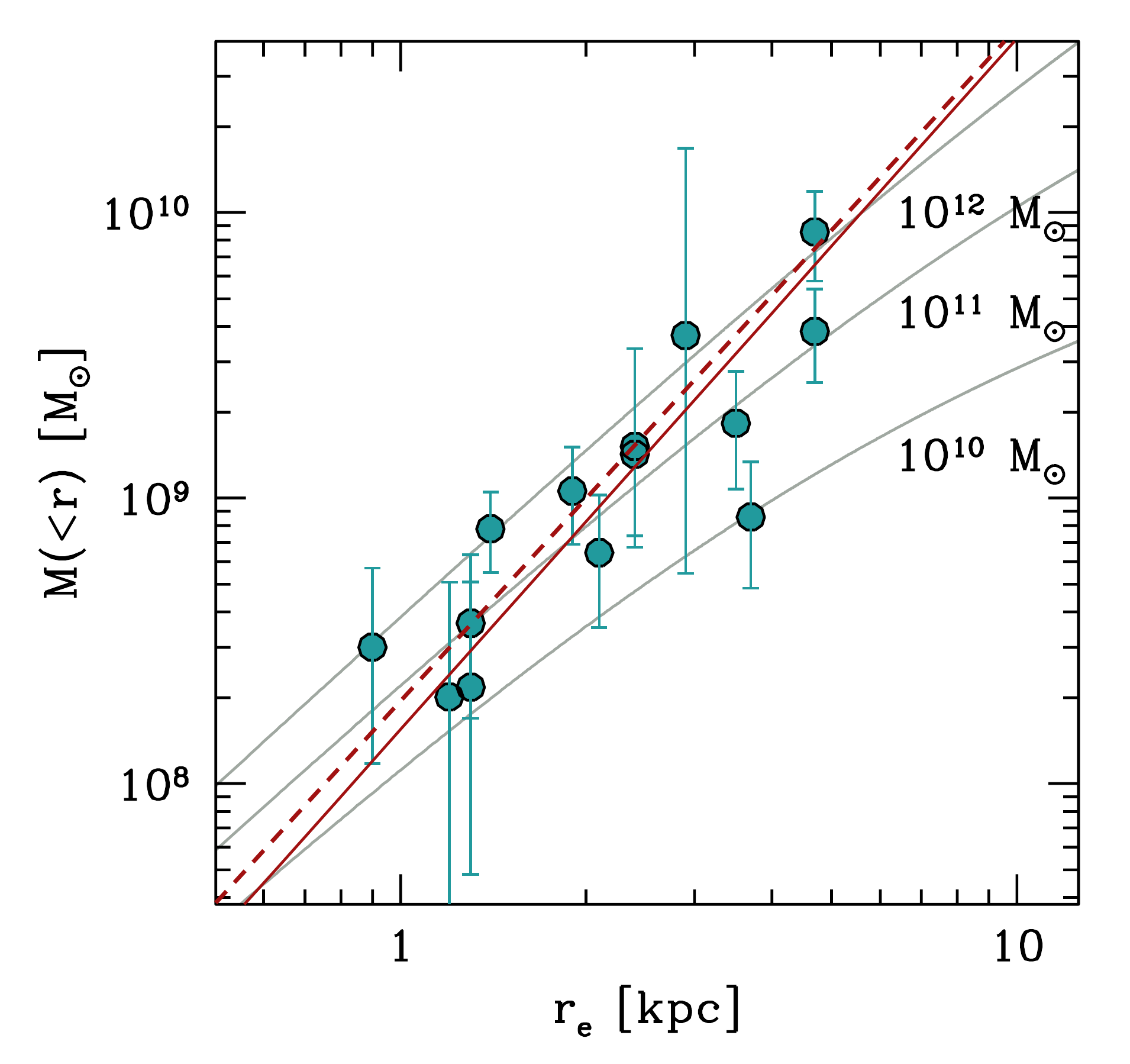}
\caption{Enclosed mass vs. half light radius for UDGs. Enclosed masses are estimated using the \cite{Walker} mass estimator. The observed UDGs are shown, with error bars reflecting only the uncertainties on the velocity dispersion (references for the UDG data are given in the text). The three curves show the enclosed mass as a function of radius for NFW halos of the given virial mass. The dotted line represents the trend derived using the FM scaling relation by \cite{Zaritsky2017}. The solid straight line represents the linear least squares fit to the data ($M(<r) = 10^{8.2}r_e^{2.4}$).}
\label{fig:scaling}
\end{figure*}

We estimate the enclosed masses with the half light radii using the estimator developed by \cite{Walker}. As shown in \cite{Zaritsky2012}, this particular mass estimator has near equivalency to the FM scaling relation used by \cite{Zaritsky2017}, facilitating our comparison. In Figure \ref{fig:scaling} we place the UDGs on the enclosed mass-radius plane for comparison to the models (NFW halos), the \cite{Zaritsky2017} result using the scaling relation (dotted line), and a standard linear least square fit to the data.

Although the data have significant scatter, the best fit line is a close match to the previous results and confirms the suggestion that as one considers UDGs with larger half light radii, one tends to systems with enclosed masses that correspond to increasingly more massive halos.
A similar, related approach has also been used by \cite{Lee2020} to estimate masses for their sample of UDGs, which included confirmation of the original relation with additional data.

Finally, we return to the issue of the two UDGs projected near NGC 1052 that we exclude. These two have small enclosed masses within their effective radii ($\sim 10^8 \text{M}_\odot$ or less), $\sim 1 - 3$ kpc depending on the adopted distance. As such, including them in the fitting results in a steeper relation, strengthening our case for a relationship between size and mass, but leading to a worse fit overall for the simple functional form we adopt.

\end{appendix}

\nocite{*}


\clearpage

\end{document}